\documentstyle[epsfig]{mn}

\newif\ifAMStwofonts

\ifoldfss
  \ifCUPmtlplainloaded \else
    \NewTextAlphabet{textbfit} {cmbxti10} {}
    \NewTextAlphabet{textbfss} {cmssbx10} {}
    \NewMathAlphabet{mathbfit} {cmbxti10} {} 
    \NewMathAlphabet{mathbfss} {cmssbx10} {} 
  \fi
  \ifAMStwofonts
    \ifCUPmtlplainloaded \else
      \NewSymbolFont{upmath} {eurm10}
      \NewSymbolFont{AMSa} {msam10}
      \NewMathSymbol{\upi}     {0}{upmath}{19}
      \NewMathSymbol{\umu}     {0}{upmath}{16}
      \NewMathSymbol{\upartial}{0}{upmath}{40}
      \NewMathSymbol{\leqslant}{3}{AMSa}{36}
      \NewMathSymbol{\geqslant}{3}{AMSa}{3E}

       \let\le=\leqslant
       \let\ge=\geqslant
    \fi
  \fi
\fi 

\ifnfssone
  \newmathalphabet{\mathit}
  \addtoversion{normal}{\mathit}{cmr}{m}{it}
  \addtoversion{bold}{\mathit}{cmr}{bx}{it}
  \newmathalphabet{\mathbfit} 
  \addtoversion{normal}{\mathbfit}{cmr}{bx}{it}
  \addtoversion{bold}{\mathbfit}{cmr}{bx}{it}
  \newmathalphabet{\mathbfss} 
  \addtoversion{normal}{\mathbfss}{cmss}{bx}{n}
  \addtoversion{bold}{\mathbfss}{cmss}{bx}{n}
  \ifAMStwofonts
    \ifCUPmtlplainloaded \else
      \UseAMStwoboldmath
      \makeatletter
      \new@mathgroup\upmath@group
      \define@mathgroup\mv@normal\upmath@group{eur}{m}{n}
      \define@mathgroup\mv@bold\upmath@group{eur}{b}{n}
      \edef\UPM{\hexnumber\upmath@group}
      \new@mathgroup\amsa@group
      \define@mathgroup\mv@normal\amsa@group{msa}{m}{n}
      \define@mathgroup\mv@bold\amsa@group{msa}{m}{n}
      \edef\AMSa{\hexnumber\amsa@group}
      \makeatother
      \mathchardef\upi="0\UPM19
      \mathchardef\umu="0\UPM16
      \mathchardef\upartial="0\UPM40
      \mathchardef\leqslant="3\AMSa36
      \mathchardef\geqslant="3\AMSa3E

       \let\le=\leqslant
       \let\ge=\geqslant
    \fi
  \fi
\fi 

\ifnfsstwo
  \DeclareMathAlphabet{\mathbfit}{OT1}{cmr}{bx}{it}
  \SetMathAlphabet\mathbfit{bold}{OT1}{cmr}{bx}{it}
  \DeclareMathAlphabet{\mathbfss}{OT1}{cmss}{bx}{n}
  \SetMathAlphabet\mathbfss{bold}{OT1}{cmss}{bx}{n}
  \ifAMStwofonts
    \ifCUPmtlplainloaded \else
      \DeclareSymbolFont{UPM}{U}{eur}{m}{n}
      \SetSymbolFont{UPM}{bold}{U}{eur}{b}{n}
      \DeclareSymbolFont{AMSa}{U}{msa}{m}{n}
      \DeclareMathSymbol{\upi}{0}{UPM}{"19}
      \DeclareMathSymbol{\umu}{0}{UPM}{"16}
      \DeclareMathSymbol{\upartial}{0}{UPM}{"40}
      \DeclareMathSymbol{\leqslant}{3}{AMSa}{"36}
      \DeclareMathSymbol{\geqslant}{3}{AMSa}{"3E}

       \let\le=\leqslant
       \let\ge=\geqslant
    \fi
  \fi
\fi 

\ifCUPmtlplainloaded \else
  \ifAMStwofonts \else 
    \def\upi{\pi}
    \def\umu{\mu}
    \def\upartial{\partial}
  \fi
\fi

\newcommand{\etal}{{et al.\ }}

\title{Simulations of deep pencil-beam redshift surveys}
\author[Yoshida et al.]
       {N. Yoshida$^1$, J. Colberg$^1$, S. D. M. White$^1$,
A. E. Evrard$^2$, T. J. MacFarland$^{1,3}$,
\newauthor H. M. P. Couchman$^4$, A. Jenkins$^5$, C. S. Frenk$^5$, F. R. Pearce$^5$
\newauthor G. Efstathiou$^6$, J. A. Peacock$^7$, and
P. A. Thomas$^8$. (The Virgo Consortium)\\
\\
$^1$Max-Planck-Institut f\"{u}r Astrophysik, Garching bei M\"{u}nchen,
D-85740, Germany\\
$^2$Department of Physics, University of Michigan, Ann Arbor,
MI-48109-1120\\
$^3$Now at 105 Lexington Avenue, Apt. 6F, New York, NY 10016\\
$^4$Department of Physics and Astronomy, McMaster University,
Hamilton, Ontario, L8S 4M1, Canada\\
$^5$Department of Physics, University of Durham, South Road, Durham,
DH1 3LE\\
$^6$Institute of Astronomy, Madingly Road, Cambridge, CB3 OHA\\
$^7$Royal Observatory, Institute of Astronomy, Edinburgh, EH9 3HJ\\
$^8$Astronomy Centre, CPES, University of Sussex, Falmer, Brighton,
BN1 9QH}
\date{Submitted to MNRAS, November 2000}

\pagerange{\pageref{firstpage}--\pageref{lastpage}}
\pubyear{2000}

\begin{document}

\maketitle

\label{firstpage}

\begin{abstract}
We create mock pencil-beam redshift surveys from very large
cosmological $N$-body simulations of two Cold Dark Matter cosmogonies, 
an Einstein-de Sitter model ($\tau$CDM) and a flat model with
$\Omega_0 =0.3$ and a cosmological constant ($\Lambda$CDM). We use these
to assess the significance of the apparent periodicity discovered by 
Broadhurst \etal (1990). Simulation particles are tagged 
as `galaxies' so as to reproduce observed present-day correlations.
They are then identified along the past light-cones of hypothetical 
observers to create mock catalogues with the geometry and the distance
distribution of the Broadhurst \etal  data. We produce 1936 (2625) 
quasi-independent catalogues from our $\tau$CDM ($\Lambda$CDM) simulation. 
A couple of large clumps in a catalogue can produce a high peak at low
wavenumbers in the corresponding 
one-dimensional power spectrum, without any apparent large-scale
periodicity in the original redshift histogram. Although the simulated 
redshift histograms frequently display regularly spaced clumps, the
spacing of these clumps varies between catalogues and there is no
`preferred' period over our many realisations. We find only a 0.72 (0.49)
per cent chance that the highest peak in the power spectrum of a $\tau$CDM 
($\Lambda$CDM) catalogue has a peak-to-noise ratio higher than that in
the Broadhurst \etal data. None of the simulated catalogues with such
high peaks shows coherently spaced clumps with a significance as high as
that of the real data. 
We conclude that in CDM universes, the
regularity on a scale of $\sim 130h^{-1}$Mpc observed by Broadhurst 
\etal has {\it a priori} probability well below $10^{-3}$.
\end{abstract}


\section{Introduction}
The redshift distribution of galaxies in the pencil-beam survey
of Broadhurst et al. (1990, hereafter BEKS) displayed a striking 
periodicity on a scale of 128$h^{-1}$Mpc. This result has attracted 
a good deal of interest over the subsequent decade, and the 
significance and nature of periodicity or regularity in the distribution 
of galaxies has remained the subject of a stimulating debate in both 
observational and theoretical cosmology.  
Although a number of studies have been devoted to the BEKS pencil-beam
survey and other similar surveys, several fundamental
questions remain unanswered. 

From the theoretical viewpoint, it is important to decide whether 
such apparently periodic galaxy distributions can occur with reasonable
probability in a Cold Dark Matter universe, or require physics beyond
the standard paradigm.
Performing large simulations can directly address this question.
The first simulation specifically designed for pencil-beam
comparisons was that of Park and Gott (1991, hereafter PG). Their
rod-shaped CDM simulation allowed them to create twelve quasi-independent
mock pencil-beam surveys similar in length to that of BEKS.
One of their samples appeared `more periodic' than the BEKS data
according to the particular statistical test they used for
comparison. Other authors (Kurki-Suonio et al.; Pierre 1990; Coles 1990; 
van de Weygaert 1991; SubbaRao \&  Szalay 1992) have used purely
geometrical models such as cubic lattices and Voronoi foams to explore
the implications of apparent regularities similar to those found by BEKS.
In particular, SubbaRao and Szalay (1992) presented a sequence of Monte
Carlo simulations of surveys of Voronoi foams, showing that such a model 
can successfully reproduce the data as judged by a variety of
statistical measures, for example, the heights, positions and signal-to-noise
ratios of the highest peaks in the power spectra. Kaiser \& Peacock
(1991) argued that the highest such peaks in the BEKS data are not sufficiently
significant to be unexpected in a CDM universe, but did not support this
conclusion with detailed simulations. Dekel et al. (1992) introduced
other statistics, more similar to those of PG, and again concluded that
the apparent periodicity seen in the real data is not particularly
unlikely in any of the toy models they used for comparison. 
Their models include Gaussian models with an extreme initial power
 spectrum with power only on scales $\sim 100h^{-1}$Mpc. They
found regular `galaxy' distributions a few per cent of the time 
and concluded that the BEKS data do {\it not} rule out
all Gaussian models.
However, these theoretical studies did not give any clear answer to the question posed above:
is the BEKS regularity compatible with the standard CDM paradigm?
We attempt to answer this below using versions of all the statistical
tools developed in earlier papers.

There have been several interesting observational developments after BEKS.
Willmer et al.  (1994) found that, if the
original BEKS deep survey at the North Galactic Pole had been carried 
out 1 degree or more to the west, many of the peaks would have been 
missed. On the other hand, Koo et al.  (1993) added new data from a 
wider survey to the original BEKS data and found the highest peak in 
the power spectrum to be further enhanced. They also analysed another 
set of deep pencil-beam surveys and found a peak of weaker
significance on the {\it same} scale, 128 $h^{-1}$Mpc. This raises
another question: is 128 $h^{-1}$Mpc a preferred length scale for 
the galaxy distribution?  Further support for such a preferred
scale has been presented by Tully et al.  (1992), Ettori et al.  (1997) 
and Einasto et al. (1997). Thus one can wonder whether a single
scale could be indicated with such apparent consistency within the
CDM paradigm.

With the important exception of the work of PG there has been 
surprisingly little comparison of the BEKS data with direct 
simulations of standard CDM cosmogonies. Even before the BEKS discovery, 
White et al.  (1987) had shown that pencil-beams drilled through periodic 
replications of their CDM simulations frequently showed a kind of 
`picket fence' regularity in their redshift distribution. 
Frenk (1991) confirmed this result and concluded that regular
patterns similar to that seen in the BEKS data are easy to find
in their simulations.
However, it is clearly dangerous to make use of periodic replications of a simulation 
when assessing the significance of apparent
periodicities in the redshift distribution. It is preferable to
simulate a volume large enough to encompass the whole survey. Furthermore,
since many independent artificial surveys are needed to establish that
the real data are highly unlikely in the cosmogony simulated, the
simulated volume must be fully three-dimensional (unlike that of PG)
to allow the creation of many quasi-independent lines-of-sight. A final
consideration is that the BEKS data reach to redshifts beyond 0.3, so 
that evolution of clustering along the survey may not be negligible.

In this paper we investigate the distribution of `galaxies' along the past
light-cones of hypothetical observers. Particle positions and velocities
on these light-cones were generated as output from the Hubble Volume 
Simulations (Evrard et al.  2000). These very large CDM $N$-body 
simulations were recently performed by the Virgo consortium and each used
$10^9$ particles to follow the evolution of the matter distribution
within cubic regions of an $\Omega=1$ $\tau$CDM ($\Omega=0.3$ $\Lambda$CDM) 
universe of side 2000 $h^{-1}$Mpc (3000 $h^{-1}$Mpc). Such large volumes
allow many independent light-cones to be generated out to
$z \sim 1$. The light-cone output automatically  
accounts for clustering evolution with redshift. The principal uncertainty
lies in how to create a `galaxy' distribution from the simulated mass
distribution. We employ Lagrangian bias schemes similar to those of
White et al.  (1987) and Cole et al.  (1998). Individual particles are
tagged as galaxies with a probability which depends only on the smoothed
{\it initial} overdensity field in their neighbourhood. The parameters of 
these schemes are adjusted so that the present-day correlations of the
simulated galaxies match observation. Many quasi-independent
mock pencil-beam surveys can
then be created adopting the geometry and the galaxy selection 
probability with distance of the BEKS surveys.

Our discussions focus primarily on the significance of the BEKS data
in comparison with our CDM samples. We begin by
following the methods used originally by BEKS, namely,
redshift counts, pair separation distributions, and the
one-dimensional power spectrum. Redshift counts are translated into
a distribution in physical distance assuming the same
cosmological parameters as BEKS. 
For the one-dimensional power spectra, the height of the highest peak is 
the most important statistic. Szalay et al. (1991) show that the
statistical significance of the highest peak of the BEKS data is at
$10^{-4}$ level, based on the formal probability for the peak height.
This calculation was disputed by Kaiser\&  Peacock (1991) because of the
difficulty in estimating the appropriate noise level. 
We calculate relative peak-to-noise ratios of 
the highest peaks in the power spectra in identical ways for real
and simulated data and so can compare the two without
needing to resolve this issue. We also apply two additional statistical 
tests for regularity, the $\Delta$ test of PG and
a `supercluster' statistic designed by Dekel et al. (1992).

Our paper is organised as follows. In section 2 we present details of 
the $N$-body simulations from which
our pencil-beam samples are drawn. In section 3 we explain our
bias scheme. In
section 4 we describe our mock pencil-beam surveys which 
mimic as closely as possible the actual observations of BEKS.
Our main results for power spectra are given in section 5.
Results are given in section 6 for the $\Delta$ test, and
in section 7 for the `supercluster' statistic. We present our 
conclusions in section 8.
 
\section{$N$-body simulation}
The simulation data we use are the so-called ``light-cone outputs'' 
produced from the Hubble Volume simulations (details are in Evrard et
al. 2001). The basic
simulation parameters are tabulated in Table 1, where $L_{\rm box}$ is the box
size in $h^{-1}$Mpc, $\Gamma$ stands
for the shape parameter of the initial power spectrum and $m_{\rm p}$ is
the mass per particle; other notations are standard.

\begin{table}
\begin{center}
\caption{Parameters of the Hubble Volume Simulations}
\begin{tabular}{cccccccc}
\hline
Model & $L_{\rm box}$ & $\Omega$ & $\Lambda$ & $h$ & $\sigma_{8}$ & $\Gamma$ &
$m_{\rm p}$ ($M_{\odot}/h$)\\
\hline
\rule{0in}{3ex}
$\tau$CDM & 2000.0 & 1.0 & 0.0 & 0.5 & 0.6 & 0.21 & 2.22$\times
10^{12}$\\
\hline
\rule{0in}{3ex}
$\Lambda$CDM & 3000.0 & 0.3 & 0.7 & 0.7 & 0.9 & 0.21 & 2.25
$\times 10^{12}$\\
\hline
\end{tabular}
\end{center}
\end{table}

\begin{figure}
\begin{center}
\vspace*{5cm}
\epsfig{file=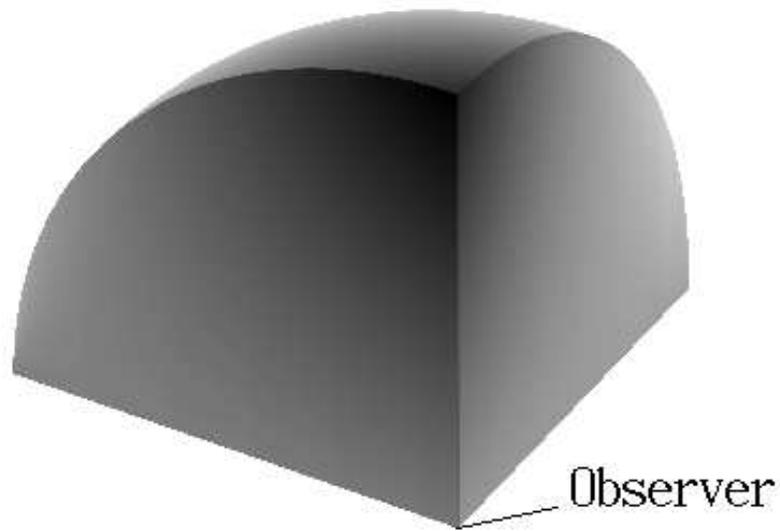,width=12cm,height=12cm} 
\caption{The shape of a light-cone and the
 observer point. The radius of the sphere is 1500 $h^{-1}$ in both 
 the $\tau$CDM and the $\Lambda$CDM models.}
\label{2octans}
\end{center}
\end{figure}

The light-cone outputs are created in the following way. 
We define an observer at a specific point in the simulation box
at the final time.  The position and velocity of each particle 
is recorded whenever it crosses the past light-cone of this observer,
and these phase-space coordinates are accumulated in a single 
data file. The evolution of clustering with lookback time (distance 
from the observer) is automatically included in such data.
As we require mock pencil-beam surveys
which reach $z \sim$0.5 (spanning $\sim$1000$h^{-1}$Mpc in physical
scale), such light-cone output is both realistic and desirable.  
We use stored data from two different light-cone outputs for each
cosmology. Each covers one octant of a sphere, and they emanate 
in opposite directions from the same point. Figure \ref{2octans} 
illustrates this geometry. In each case we use data out to a
comoving distance of 1500$h^{-1}$Mpc, corresponding to redshift
0.77 in the $\tau$CDM model and $0.58$ in the $\Lambda$CDM model.
For the $\tau$CDM case the total length covered is larger than
the side of the simulation box, but this has a negligible effect
on the mock BEKS surveys we construct. 

\section{Galaxy selection}
To create realistic mock surveys we have to select particles as
galaxies with the same distribution in depth as the real data and with
an appropriately `biased' distribution relative to the dark matter.
We do this in two stages. First we identify a biased subset of the 
particles chosen according to the value of the smoothed linear
mass overdensity at their position at high redshift. The parameters
defining this identification are chosen so that the two-point
correlation function of the identified `galaxies' at $z=0$ matches
the observed correlation function of low redshift galaxies.
For the $\tau$CDM model, we are able to achieve this while  
retaining about two-thirds of the simulation particles as `galaxies'. 
The resulting comoving `galaxy' number density is 0.08
$h^{3}$Mpc$^{-3}$.  For the $\Lambda$CDM model we get a number 
density of `galaxies' in the range 0.02 to 0.033 $h^{3}$Mpc$^{-3}$ 
depending on the bias scheme. This lower number density is due to 
the low number density of the dark-matter particles in this model. 
The second stage is to mimic the effect of the apparent magnitude limits
of the real galaxy surveys by including `galaxy' particles into the
final mock catalogues with a probability which depends on distance
from the observer. Since this stage is independent of the first,
we are effectively assuming that the clustering of galaxies is
independent of their luminosity. Our radial selection function is based
on those directly estimated for the BEKS surveys.

\subsection{Lagrangian Bias}

Cole et al. (1998) developed and tested a set of 
bias schemes to extract `galaxies' from $N$-body simulations.
The procedure we use for the first stage of our galaxy selection
is similar to their Model 1, but has
a different functional form for the probability function.
Since we need a bias factor greater than unity for the $\tau$CDM model 
and less than unity for the $\Lambda$CDM model (a result of the 
differing mass correlations in the two cases) `galaxies' need to avoid 
regions of low initial density in the $\tau$CDM simulation and to avoid 
regions of high initial density in the $\Lambda$CDM simulation.
We begin by smoothing the density field at an early time with a Gaussian, 
$\exp(-r^{2}/2 r_o^{2})$ with $r_o=3 h^{-1}$Mpc and assigning the corresponding
overdensity $\delta$ to each dark-matter particle.
Then a normalised overdensity $\nu=\delta/\sigma_{s}$ is computed,
where $\sigma_{s}$ is the root mean square value of particle $\delta$-values.
Finally, we define a probability function $P(\nu)$ which determines
whether a particle is tagged as a `galaxy'. We 
random-sample dark-matter particles for tagging as galaxies based
on this probability. Once tagged as a galaxy in this way, particles
remain tagged throughout the simulation, and so become potentially visible
in our mock surveys whenever their world-line crosses a light-cone.

\begin{figure}
\begin{center}
\epsfig{file=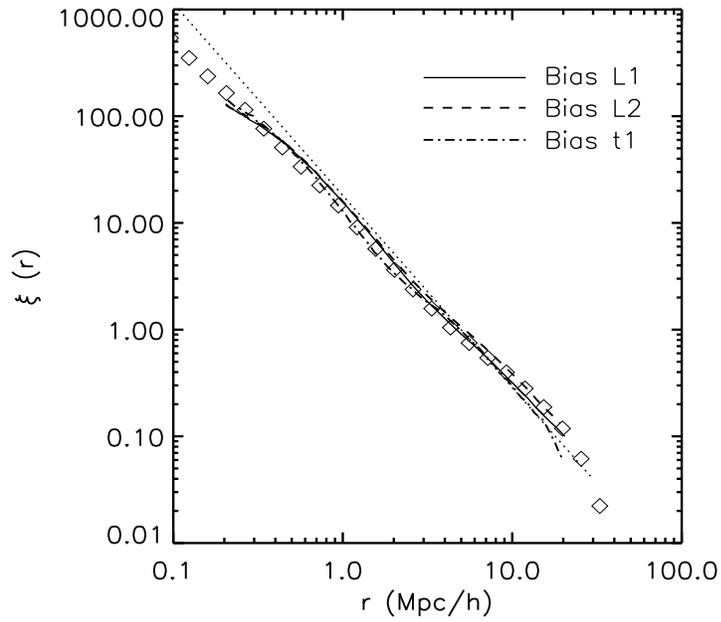,width=9.5cm,height=9cm} 
\caption{The two-point correlation functions of the biased
`galaxies'. The dash-dotted line is the `galaxy-galaxy'
correlation of our model bias {\bf t1} and the solid line is for
bias {\bf L1}, the dashed line for bias {\bf L2}.
The open squares are the observational data from the 
APM survey (Baugh 1996) and the dotted line is the assumed galaxy correlation function in Szalay
et al. (1991) for the analysis of the BEKS survey. 
The curves fit well both the APM data and the Szalay et al. model.}
\label{corr2}
\end{center}
\end{figure}

For all the bias models described below the `galaxy-galaxy' correlation
function 
was calculated in real space within a cubic box of side 200 $h^{-1}$Mpc 
with the observer at one corner. These correlation functions are shown
in Figure \ref{corr2}.

{\bf $\tau$CDM model bias t1:} For the $\tau$CDM model, we chose 
a simple power law form $P(\nu)\propto  (\nu-\nu_{c})^{0.2}$ for the
probability function. We impose a threshold at $\nu=\nu_{c}=-0.55$ 
below which the probability is set to be zero. This suppresses the 
formation of `galaxies' in voids. 
These parameters were determined by matching the present-day two-point
correlation function of the `galaxies' to the
observational result for the APM survey (Baugh 1996, see Figure \ref{corr2}) on
length scales from 0.2 $h^{-1}$Mpc to 20$h^{-1}$Mpc. We note here
that in our $N$-body simulations the gravitational softening length
is 0.1$h^{-1}$Mpc.

{\bf $\Lambda$CDM bias model L1:}
For the $\Lambda$CDM model, we must `anti-bias' because the predicted
mass correlations on small scales are substantially larger than
observed galaxy correlations (see, for example, Jenkins et al. 1998).
We set a sharp {\it upper} cut-off at $\nu_{c}=1.34$, 
above which $P(\nu)$ is zero. All particles below this threshold
are equally likely to be `galaxies' ($P=$ const).
Although this may seem unphysical, more realistic
modelling of galaxy formation in $\Lambda$CDM models does indeed
produce the anti-bias required for consistency with
observation, albeit through a more complex interplay of statistical 
factors (Kauffmann et al. 1999; Benson et al. 2000). We use a simpler 
scheme in order to produce the desired two-point correlation 
function; on scales of interest here, only a small anti-bias is 
necessary.

{\bf $\Lambda$CDM bias model L2 :}
For comparison purposes, we applied a second bias model to the 
$\Lambda$CDM simulation. We set an additional {\it lower} threshold at 
$\nu_{low}=-0.7$ below which we again set $P=0$. Thus the probability 
takes a non-zero (and constant) value only in the range
$\nu_{low} \le \nu \le \nu_{high}$, where now $\nu_{high}=0.9$.  
This model fits the observed correlations of galaxies just as well
as {\bf L1} but enhances the emptiness of voids \\

\begin{figure}
\begin{center}
\epsfig{file=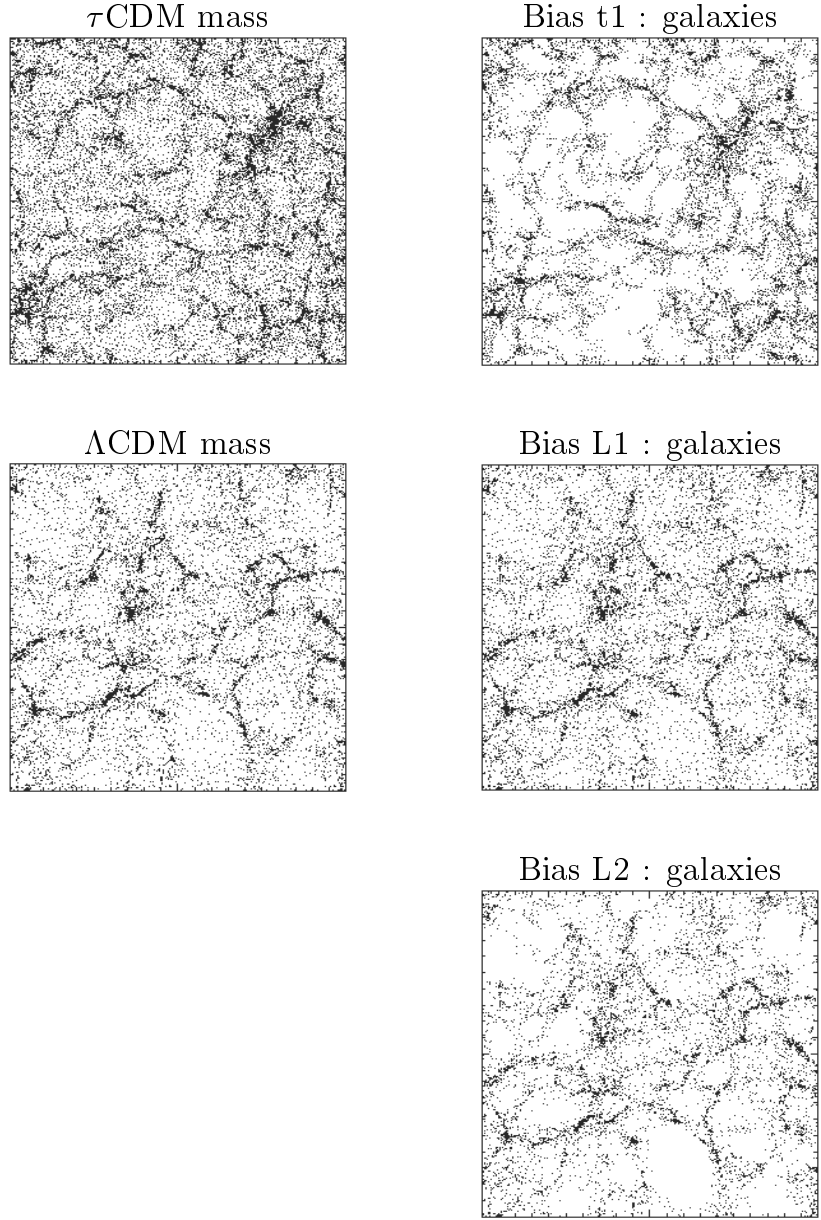,width=12cm,height=18cm} 
\caption{The distribution of the simulation dark-matter
 particles(left panels) and the biased `galaxies'(right
panels) in real space. 
 The panels show $200 \times 200 \times 10$ $h^{-3}$ Mpc$^{3}$
 slabs. Note the difference in the number density of the simulation 
dark-matter particles between the $\tau$CDM model
 (0.125$h^{3}$Mpc$^{-3}$) and the $\Lambda$CDM model (0.037$h^{3}$Mpc$^{-3}$).
Strong-bias effects are apparent in {\bf t1} and {\bf L2} whereas in
 {\bf L1} essentially no bias is seen.}
\label{double}
\end{center}
\end{figure}

Figure \ref{double} illustrates bias effects by comparing
the distribution of dark-matter particles and of `galaxies' 
in a thin slice through part of the simulation box at $z=0$. 
For model {\bf L1} the effect is difficult to detect visually,
whereas the effect of the 
lower cut-off in model {\bf L2} is obvious.
Similarly, for the $\tau$CDM model, 
underdense regions(voids) are clearly accentuated in the `galaxy'
distribution relative to the dark matter distribution.
Cole et al. (1998) show similar plots to demonstrate how strong
bias in high density model universes maps underdense regions in 
the mass distribution onto voids in the galaxy distribution. 
Such contrasted voids are generally seen in strongly biased 
models regardless of the functional form of $P(\nu)$.

\begin{figure}
\begin{center}
\epsfig{file=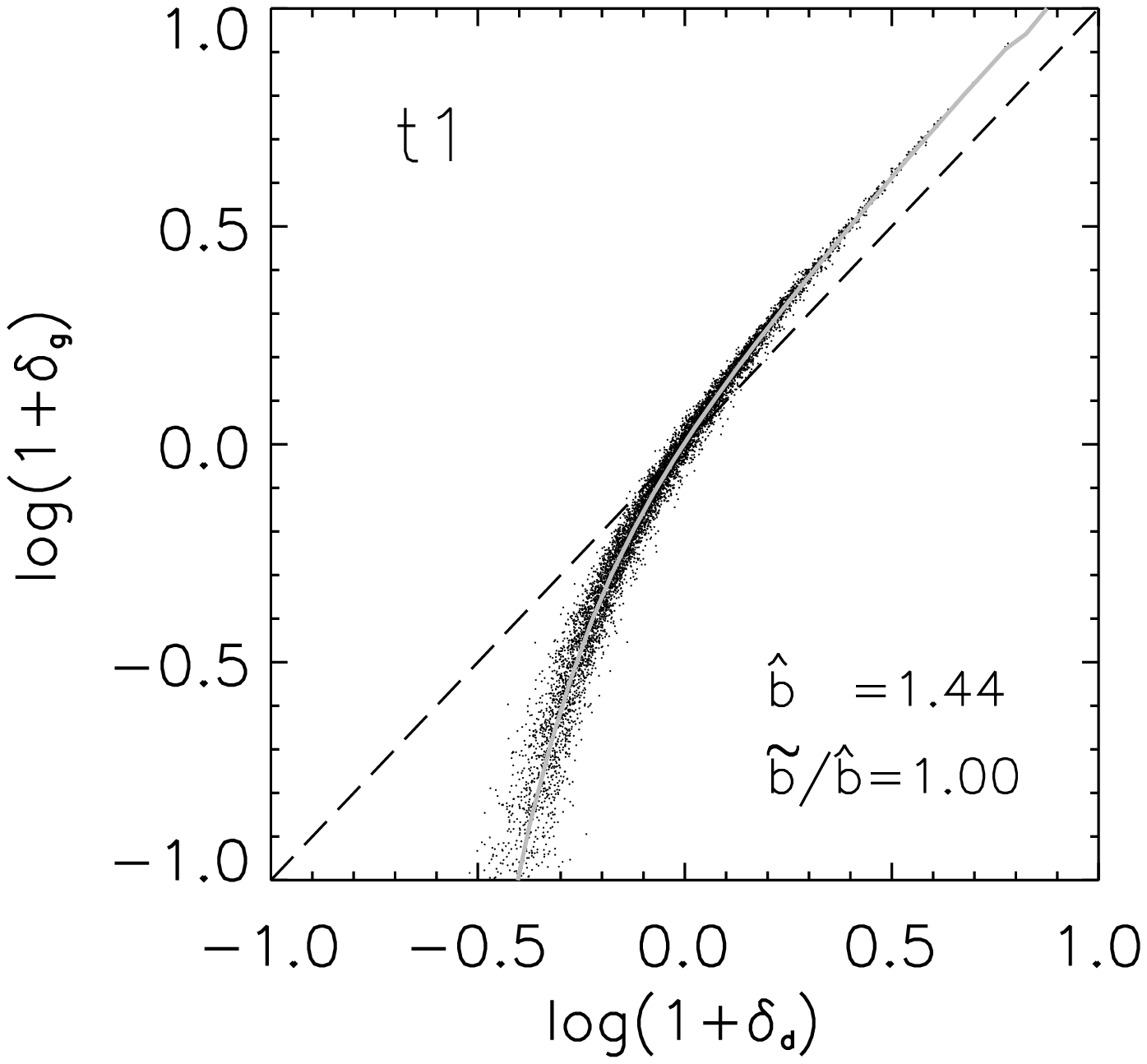,width=8cm,height=6cm}\\
\epsfig{file=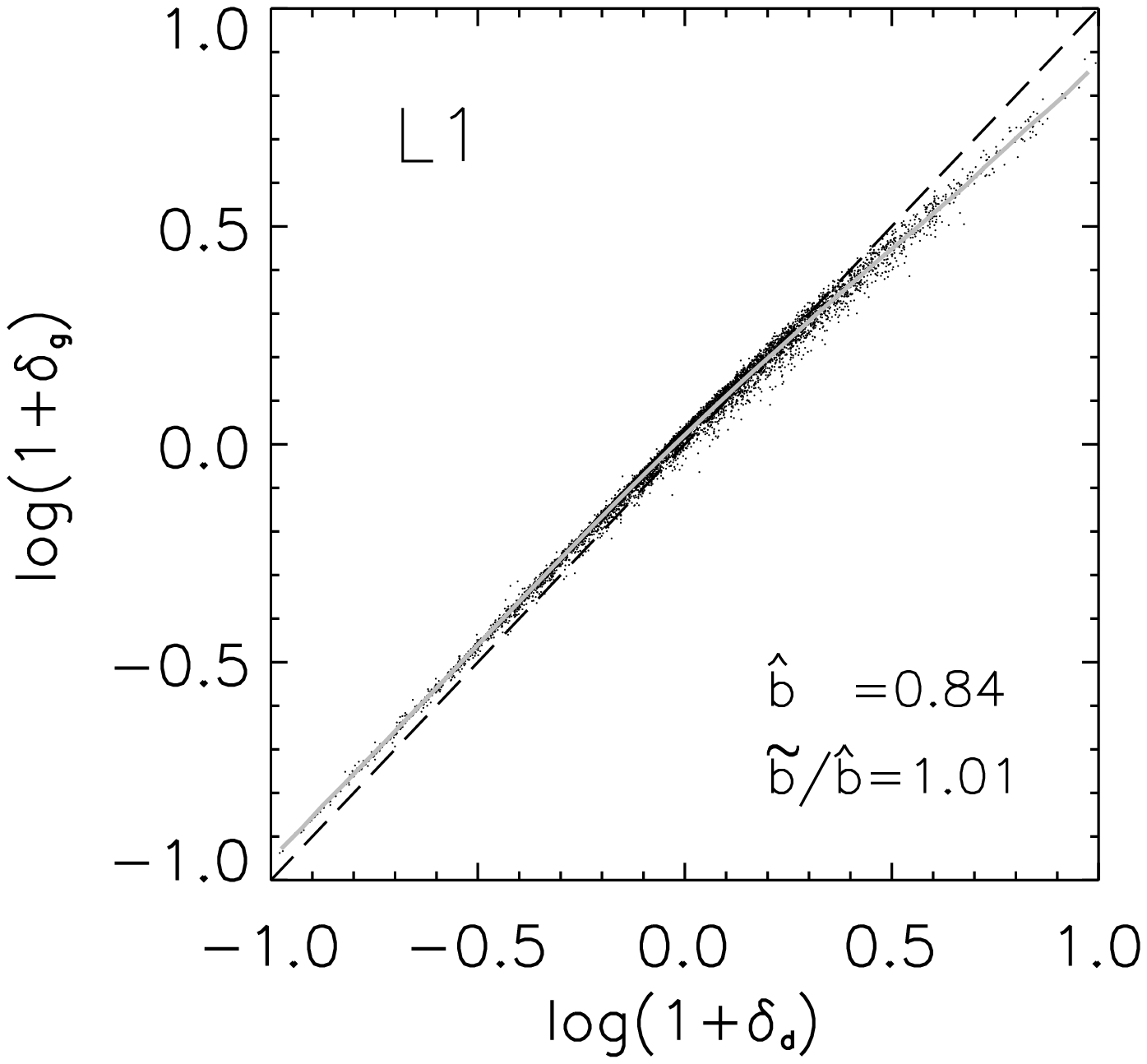,width=8cm,height=6cm}\\ 
\epsfig{file=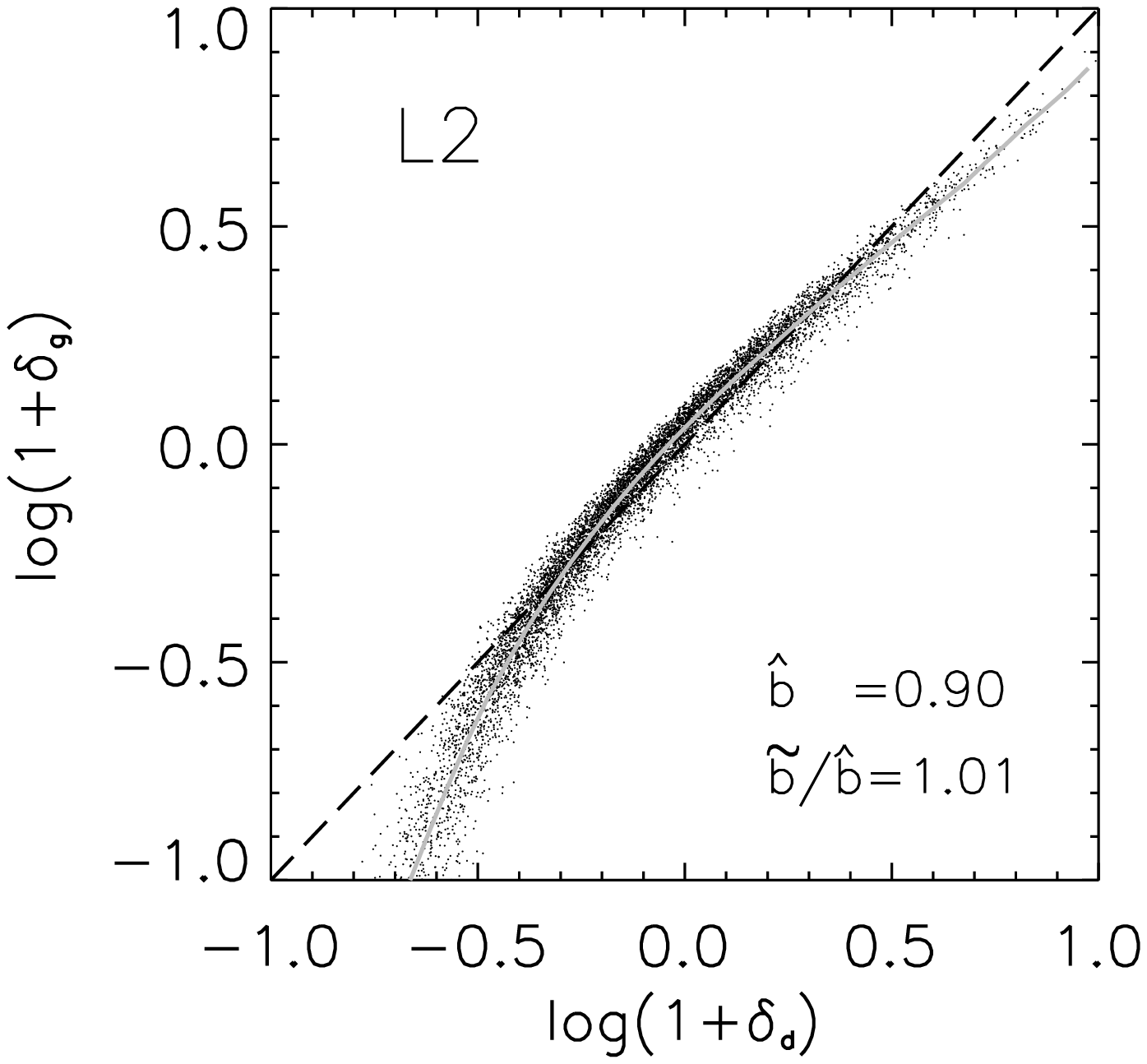,width=8cm,height=6cm} 
\caption{The joint distribution of the overdensity fields of
`galaxies' and mass, both smoothed with a 8$h^{-1}$Mpc tophat window.
The grey lines show the mean biasing function $b(\delta)$. Dashed
lines show a linear relation $\delta_{g}$ = $\delta_{d}$ for a
reference. The measured biasing parameters $\hat{b}$ and 
$\tilde{b}/\hat{b}$ are given in each panels.}
\label{biasfunc}
\end{center}
\end{figure}

 We can quantitatively study the bias in our models by measuring
the nonlinear biasing parameters introduced by Dekel \& Lahav (1999) 
(see also Sigad et al. 2000; Somerville et al. 2000).
We compute the slope $\hat{b}$ and nonlinearity $\tilde{b}$
following the procedure described in Somerville et al. (2000).
Figure \ref{biasfunc} show the biasing relation between the `galaxy' density field
$\delta_{g}$ and the dark matter density field $\delta_{d}$,
smoothed with a 8$h^{-1}$Mpc scale top-hat filter. For each of our
bias models, the mean biasing function $b(\delta_{d})$, and its moments
\begin{equation}
\hat{b}=\frac{<b(\delta_{d})\delta_{d}^{2}>}{\sigma^{2}},\;\;\;
\tilde{b}^{2}=\frac{<b^{2}(\delta_{d})\delta_{d}^{2}>}{\sigma^{2}}, 
\end{equation} 
are given in Figure \ref{biasfunc}. In the above expression we have used
$\sigma^{2}=<\delta_{d}^{2}>$ for the standard deviation.
Strong biasing in {\bf t1} and anti-biasing (for $1+\delta_{d} > 0$)
in {\bf L1} and {\bf L2} are clearly seen, and reflected in the values
for the effective {\it slope}; $\hat{b}=1.44$ for {\bf t1},
$0.84$ for {\bf L1} and $0.90$ for {\bf L2}.

\section{Survey Strategy} 
\subsection{Geometry}
We construct artificial surveys with a geometry very similar
to that of the data analysed by BEKS. This consists of
four surveys -- a deep and a shallow survey near each Galactic Pole. 
The northern deep survey lies within a cone of 40-arcmin
diameter about the pole and is made up of a set of roughly circular 
patches each 5-arcmin in diameter. About 10 small patches were surveyed 
but not all were completed by the time of writing so that the exact 
number of patches used in BEKS is unclear.
For our artificial surveys we choose 9 circular patches within the 
40-arcmin diameter cone, each of diameter 5 arcminutes. We place these
irregularly and ensure no overlaps between them. 
For model {\bf L2} the number density
of `galaxies' is too small to match the observations, so
we increased the diameter of our patches to 7-arcmin. 
Although this widening results in a slight increase in the effective
survey volume, the small patches still lie well within the larger cone of
40-arcmin diameter. The volume increase is compensated in the radial
selection we decribe below, so that the resulting `galaxy' distribution
is consistent with the desired distribution given in BEKS.
For the deep-narrow pencil-beams, the transverse length scale is much 
smaller (the cone diameter is $\sim 4h^{-1}$Mpc at z=0.2 where the
radial selection function takes its maximum value) than the $100h^{-1}$Mpc
scale we address, so the increase in the patch width does not affect our results. 
In all cases
the redshift counts in all patches were binned together to create a
single deep survey. The northern shallow survey has a simpler geometry.
A square area of about 14 square degrees is selected near the Galactic Pole, 
but with its centre offset by 7 degrees. The
magnitude limit of the shallow survey is about 5 magnitudes brighter 
than that of the deep survey.

Towards the South Galactic Pole both surveys are centred very close
to the pole itself. The deep survey is confined within a cone of 
20-arcmin diameter,while the shallow survey covers an area of 14
square degrees and has a magnitude limit about 4 magnitudes brighter.

When making an artificial survey we choose a random direction
in the simulation as the Galactic polar axis and then define all
areas on the artificial sky with reference to this direction. The 
light-cone outputs from our Hubble Volume simulations cover enough
`sky' to allow us to make well over 1000 near-independent artificial
surveys.

\subsection{Radial selection}
`Galaxies' projected in our survey regions are assigned weights for 
selection depending on their distances. We use 
the estimated galaxy distributions given in Figure 1 of BEKS to
define the relevant selection functions for each survey. 
The data are read off in redshift bins of width $\Delta z=0.005$ 
for the deep surveys and $\Delta z=0.001$ for the shallow surveys.
We then derive a smoothed model galaxy distribution $dN/dz$ for each survey
and compute the corresponding comoving number densities 
from the number counts and the volume
elements given by the survey geometry and the assumed cosmology. 
As explained above, we had to increase the size of the patches in
the northern deep survey in case {\bf L2} in order to get the
correct mean counts. This is easily accounted for by appropriate
renormalisation.
These radial selection functions are used as sampling 
probabilities to determine whether a particular `galaxy' is included 
in a catalogue or not. 
We normalise our probabilities by
matching the mean number of `galaxies' in each survey
to the number of galaxies in BEKS data. This matching is done for the
4 surveys independently. For consistency, the normalisation coefficients 
obtained are then kept constant when constructing all realisations
for a particular model. 

\subsection{Peculiar velocities}
The peculiar velocities of `galaxies' must be taken into account 
to create realistic mock redshift surveys. 
We simply assign our `galaxies' the peculiar velocities of their 
corresponding dark-matter particles. Thus, while the spatial 
distribution of `galaxies' is biased, there is no additional
bias associated with their peculiar velocities.  
On small scales peculiar velocities lead to `finger-of-God' effects which
suppress power in the apparent spatial distribution at high 
wavenumber. 
In our mock catalogues the root mean square values of the
`galaxy' line-of-sight peculiar velocities  
are 342 km/sec in {\bf t1}, 358 km/sec for {\bf L1} and {\bf L2}.
The redshift bin width shown in BEKS is 
$\Delta z$=0.005 for the deep surveys, which translates $\sim$1500 km/sec
in recession velocity. Therefore, the assigned peculiar velocities of
`galaxies' do not smear out the {\it true} width of clumps 
in one-dimensional distributions, while they reflect properly
the underlying velocity field.
At the intermediate and small wavenumbers
corresponding to the linear and quasi-linear regime, 
they increase the apparent power (e.g. Kaiser and Peacock 1991). 
These line-of-sight distortions reflect the enhanced contrast produced by infall onto 
superclusters. 
It is thus important to include the peculiar
velocities when comparing simulations to the structures seen in the
BEKS data.

\section{Galaxy distribution and power spectrum analysis}
The geometry of our light-cone datasets allows the axis of
our artificial surveys to lie anywhere within one octant
of the `sky' (see Figure \ref{2octans}). We construct an ensemble of mock
surveys with axes distributed uniformly across this octant
in such a way that the areas covered by the corresponding
deep surveys do not overlap. We end up with 1936 quasi-independent
deep surveys for our $\tau$CDM model. 
For the $\Lambda$CDM case an additional pair of
light-cone outputs were stored, allowing us to construct 2625
disjoint deep surveys.
To these deep pencil beams we add shallow surveys, 
whose volumes then have slight overlaps with those of neighbouring 
surveys. In practice, however, rather few `galaxies' appear in
more than one of our mock catalogues.

A series of plots of the redshift distribution and derived
statistics are given for selected `mock BEKS surveys'  
in Figure 6, which consists of 6 sets of 3 figures. These can be
compared with Figure 5, which is actually
for the real BEKS data, which we reproduce here for comparison
with our simulation results. We read these data from Figure 2 in 
Szalay et al. (1991) where they are given as a histogram of
bin width 10 $h^{-1}$Mpc; when necessary for our analysis, 
we assume that the galaxies in each bin are uniformly distributed
across the bin. The particular mock surveys in the following 6
plots were chosen to illustrate a variety of points made in
the following sections. 

\subsection{One-dimensional distribution}
In each set of plots in Figure 6, the top panel shows the
distance histogram of `galaxies' in the combined
deep and shallow surveys. 
The total number of galaxies in these combined surveys is
given in this panel. 
We have assumed an Einstein-de Sitter universe
for both of our models when converting redshift to physical distance,
although the actual value of $\Omega$ is 0.3 in the $\Lambda$CDM case. 
This apparent inconsistency is needed to allow a direct comparison 
with the analysis in BEKS where $\Omega=1$ was also assumed. 
Szalay et al. (1991) noted that using low values of $\Omega$ 
to convert redshift to distance reduces the significance of the
apparent periodicity in the BEKS data. Throughout this paper we assume 
$\Omega=1$ for this conversion. 

In Figure 6, panel (a) shows one of the best catalogues in our {\bf t1}
ensemble in that it gives the impression that `galaxies' are distributed 
periodically and, in addition, the 1-D power spectrum shows a sharp and high 
peak. Panel (b) shows the same features but with a smaller characteristic
spacing. In each of these plots we mark the best periodic
representation of the data in the same way as BEKS. 
We determine the characteristic
spacing from the position of the highest peak in the power spectrum,
and we adjust the phase to match the positions of as many big
clumps as possible. 
The characteristic spacing is indicated by the vertical dashed lines in
the top panels in Figure 6. 
Panel (c) shows a good example whose power spectrum has a very high peak
while the actual distance distribution does not show a periodic feature
(discussed in section 5.3). Panels (d) and (e) show the best examples
from our model {\bf L1} and {\bf L2}, respectively, which show a good
visual impression that `galaxies' are spaced regularly. Finally
panel (f) shows an example from model {\bf L2}, which has a large characteristic 
length scale of $\sim 200h^{-1}$Mpc. 

\subsection{Pairwise separation histograms}
From the apparent distance distributions of the `galaxies', it is
easy to produce histograms of pairwise distance differences which
can be used to search for characteristic scales in the structure
within our pencil-beam surveys. Such pair counts
are shown in the middle plot of each panel in Figure 6. 
These counts typically display a series of peaks and valleys
which are particularly prominent in panels
(a), (e) and (f), and, as noted by BEKS themselves, in the original
BEKS data. Notice that these peaks appear regularly spaced
as indicated by the dashed lines in these panels. 
The contrast between peaks and valleys can be used as a measure of
the strength of the regularity. For the BEKS data the height difference 
between the first peak and the first valley is about a factor of 3, while 
the corresponding numbers are 2.4, 2.2 and 3.4 in panels (a), (e), and
(f), respectively. Many of our artificial samples show a more complex
behaviour, however. In panel (b) there is a deep valley at 
150$h^{-1}$Mpc followed by a high peak at 200$h^{-1}$Mpc; the contrast
is a factor of 5.3 despite this uneven spacing. A robust and intuitive
definition of contrast is difficult to find.
An alternation of small-scale peaks and valleys can coexist with
apparently significant larger scale variations as is clearly seen
in panel (c). If we focus specifically on the strongest peaks and valleys,
their ratio, and indeed even their identification can depend on
the specific binning chosen for the histograms. Because of these
ambiguities we will not use these distributions further for
quantitative analysis in this paper.

\subsection{Power spectra}
In order to compare our results directly with BEKS we calculate
one-dimensional power spectra for our samples using the
method described in Szalay \etal (1991). 
Each galaxy is represented by a Dirac delta-function at the distance
inferred from its redshift (including its peculiar velocity). The
power in each Fourier component is then
\begin{equation}
f_{k}=\frac{1}{N_g}\sum_{n}\exp(2\pi i k r_{n}),
\end{equation}
\begin{equation}
P_{k}=|f_{k}|^{2}
\end{equation}
where $N_g$ is the total number of galaxies in the sample.
The power spectra calculated in this way
for our various samples are shown in
the bottom plots of each panel in Figure 6. Our units are such that the
wavelength corresponding to wavenumber $k$ is 1000/$k$
$h^{-1}$Mpc. In panels (a) and (b) visual impression of the 
separation of clumps is consistent 
with the wavelength inferred from the power spectra. 
The highest peak is at $k$=8.0 in panel (a) and $k=$16.0
in panel (b), giving wavelengths of 125$h^{-1}$Mpc
and 62.5$h^{-1}$Mpc, respectively. As we shall see, 
there is no unique length scale inferred from the power spectra.

If a pencil-beam penetrates a rich cluster, 
an interesting feature can arise. For example, in panel (c) 
there is a single large
cluster at 600 $h^{-1}$Mpc. Together with a few other clumps 
of moderate size, it produces a
very high peak in the power spectrum {\it
without} the distance distribution as a whole giving a visual 
impression of regularity (c.f. the top panel of panel (c)).
Many of our samples in both the $\tau$CDM and $\Lambda$CDM
ensembles show high peaks in the power spectra with no apparent periodicity.
Thus a very high peak in the power
spectrum, particularly at low wave-number, is a poor indicator 
of the kind of regularity which is so striking in the original
BEKS data. Interestingly, as Bahcall (1991) pointed out, if one of the BEKS survey beams
passed near the centre of a rich cluster, the galaxy count in
the corresponding distance bin would have been much larger than the maximum
of 22 seen in the actual BEKS data (see also Willmer et al.  1994). 
(For comparison, the maximum bin count in the histogram of panel (a) 
is 23.) We note that the Poisson sampling noise in each power spectrum can
be estimated as $1/N_{g}$. As a result, a big clump 
raises the statistical significance of
`structure' both by enhancing the strength of peaks and
by lowering the estimated noise.

In order to compare samples with different total numbers of
`galaxies', we calculate the signal-to-noise ratio of the highest peak 
in the power spectrum following the procedure of Szalay et al. (1991). 
We define the peak-to-noise ratio of a sample as, $X$=(peak
height)/(noise level) where the noise is estimated from the sum
in quadrature of the Poisson sampling noise and the clustering noise,
\begin{equation}
f_{0}=\frac{1}{N_{g}}+\frac{\xi_{0}}{M},
\end{equation}  
where, as before, $N_{g}$ is the total number of galaxies in the
sample, $\xi_{0}$ is the small-scale two point correlation function averaged over a cell of 
depth 30$h^{-1}$Mpc, and $M$ is the number of cells along the survey axis.
We use the value $\xi_{0}/M = 1/80$ as in Szalay et al. (1991).
Although Szalay et al. derived this formula from a simple model
with cylindrical geometry, they showed that this
estimator agrees well with another internal estimator 
calculated from the cumulative distribution of power. 
To facilitate direct comparison with the earlier work
we also use equation (3) to compute signal-to-noise ratios for the
highest peaks in our samples. These S/N ratios are given in
each of the power spectrum plots in Figure 6. 
Figure \ref{c_d} shows both the differential and the cumulative
distribution of peak-to-noise ratio in our mock surveys. The principal
difference between the $\tau$CDM 
and $\Lambda$CDM ensembles lies in the position of the peak in the
differential
count. For both {\bf L1} and {\bf L2} the peak is at smaller $X$ 
than in {\bf t1}.
This difference can be traced to the value of $\Omega$ we assume for analysis.
Adopting $\Omega=1$ for converting redshift to physical distance
causes the value of $M$, the number of cells of width
30$h^{-1}$Mpc
along the survey axis, to be underestimated for $\Lambda$CDM.
Using the noise estimator (3) with this
value of $M$ then overestimates the noise levels for {\bf L1} and {\bf L2}
(see SubbaRao and Szalay (1992) for discussion of a similar point).

We plot in Figure \ref{peaks} the wavenumber distribution of the peaks whose
S/N ratios are higher than that of the original BEKS data ($X$=11.8). 
We find 14 samples satisfy this condition in {\bf t1}, 7 in {\bf L1} and 
13 in {\bf L2}. The distributions of the peaks with $X > 8.0$
(the `tails' of the number count in Figure \ref{c_d})
are also shown in Figure \ref{peaks}. 
 By checking the distance distributions
we have found that highly significant peaks at
$k \le 5$ are almost always due to one or two strong clumps, as
noted above. Very few catalogues give a high peak on scales
similar to the BEKS data. It is interesting that the frequency of such
catalogues is significantly higher in {\bf t1} than in {\bf L1} and {\bf L2}.
The difference is primarily due to the number density of rich clusters
over the redshift range surveyed. Deep pencil-beams in our {\bf L1} and {\bf L2}
models have more chance than in {\bf t1} to hit a rich cluster at
redshift $\sim 0.3 - 0.5$. Then high peaks in the power spectra tend to appear 
on small wavenumbers in {\bf L1} and {\bf L2} for the reason explained
above.

Overall, we conclude that although roughly half a per cent of our
mock surveys give a power spectrum peak stronger than that of the
BEKS data, very few of these actually correspond to redshift 
distributions with similar regularity and similar spacing of the
spikes. We now study this further by considering two additional
tests for regularity which have been used on the BEKS sample. 
 
\section{PG $\Delta$-test}

In comparing their own simulation to the BEKS data, Park \& Gott
(1991) made use of a test specifically designed to probe the apparent
``phase-coherence'' of the series of redshift spikes.
For each `galaxy' they calculated the distance 
to the nearest tooth of a perfectly regular comb-like 
template. They then ratioed the mean of this distance to the
separation of the teeth, and minimised the result over the period and
phase of the template. Let us call the resulting statistic $\Delta$.
Then a distribution in which each galaxy is at some node of a regular
grid will give $\Delta=0$, and a uniform distribution in distance
would give $\Delta=0.25$ in the large-sample limit. In our application
of this test we restrict the range of possible periods to 50 -- 500 
$h^{-1}$ Mpc. For the BEKS data we obtain $\Delta$=0.165 for a best
period  of 130$h^{-1}$ Mpc. Our value of $\Delta$ differs from that
given by PG because they applied the test only to the deep
surveys while we use the combined deep and shallow BEKS data.
Among the 1936 samples in our {\bf t1} ensemble, 
209 have lower values of $\Delta$ than the BEKS data; for
the {\bf L1} and {\bf L2} ensembles the corresponding numbers
are 134/2625 and 127/2625 respectively. According to this test,
therefore, the observed sample appears only marginally more regular than
expected in our CDM cosmologies.

Within each of our ensembles the significance of the regularity in
the BEKS data appears somewhat higher than was estimated by PG.
Their simulation ensemble was made up of only 12 mock catalogues
of which one had lower $\Delta$ than the BEKS deep data. The
median $\Delta$ for these twelve was 0.1695, while we find
medians of 0.176, 0.180 and 0.189 for the combined deep and
shallow data in ensembles {\bf t1}, {\bf L1} and {\bf L2} respectively.
The difference with PG is probably small enough to be attributed to
the small number of realisations in their ensemble.
Figure \ref{lowdelta} shows the distribution of periods for
catalogues in each of our own ensembles 
with lower $\Delta$s than the BEKS data. It is interesting that
relatively small periods are favoured and that there is no
preference for values in the range $120-130h^{-1}$Mpc.
Ettori et al. (1997) used a related test, the comb-template test
(Duari et al. 1992), to analyse four pencil-beam surveys near the
South Galactic Pole. They found a best period near 130
$h^{-1}$Mpc in two of these four directions, in apparent agreement
with the BEKS result.

In summary, the difference in regularity between the BEKS sample
and our CDM mock-catalogues is less significant when
measured by the $\Delta$-test than when measured using the
power spectrum test of the last section. Nevertheless, for periods
near $125 h^{-1}$Mpc there are few CDM samples which are more
regular than the BEKS data.
In addition, our result appears insensitive to the choice of biasing; 
we find essentially no difference
between {\bf L1} and {\bf L2} in Figure \ref{lowdelta}. 
This is puzzling since Figure \ref{double} shows clear differences in 
the emptiness of the voids in the two cases. Apparently the value of
$\Delta$ is more sensitive to departures from regularity in 
the spacing of the walls than it is to the density contrast of the
voids. 
 
\section{Supercluster statistics}
Dekel et al.  (1992) proposed an alternative technique for 
assessing apparent periodicity in data samples like
that of BEKS. In this section we use the term `supercluster' to refer
to clumps in the one-dimensional redshift histograms
derived from such pencil-beam surveys, even though these do not
correspond precisely to the superclusters (or walls or filaments) 
seen in fully three-dimensional surveys. The method of Dekel et al.
is based on the redshift distribution of supercluster centres 
rather than on that of individual galaxies. The first step is to 
correct the galaxy redshift histogram for the 
survey selection function.  We do this by weighting each galaxy
by the inverse of the selection function for the particular
survey of which it is a part (North or South, shallow or deep).
This reverses the  procedure by which we created our mock catalogues
from the simulations. To avoid overly large sampling noise
where the selection function is small, we restrict our
redshift histograms to $z\le0.31$ for the SGP survey and $z\le0.5$ 
for the NGP survey (see Dekel et al. 1992). We smooth these
histograms with a Gaussian of variance $l^2$ and identify 
supercluster centres as local maxima of the result. (Note that,
following Dekel et al., no threshold is applied.)
We have tried smoothing lengths $l$ between 
20 and 40 $h^{-1}$Mpc, but find our results to be insensitive to
the exact value within this range. In what follows we set $l$ =25
$h^{-1}$Mpc.

Given a distribution of the supercluster centres, the characteristic 
period is determined in the following way. As a first estimate we
take the mean separation $L_{m}$ between neighbouring superclusters. 
Next we apply the PG $\Delta$-test for periods $p \in 
[0.5L_{m},2.0L_{m}]$. The value of $p$ in this range which minimises
$\Delta$ is taken as
the characteristic period of the distribution. For this period
we calculate the Rayleigh statistic $R$ as follows
(Dekel et al. (1992) and Feller(1971)). The positions of the
supercluster centres are mapped onto a circle of circumference
$p$. Consider the $n$ unit vectors ${\bf u}_i$ which point from 
the centre of the
circle towards each of the $n$ superclusters. Denote their 
vector average by $\langle{\bf u}\rangle$, the modulus of
$\langle{\bf u}\rangle$ by $V$ and define $R=1-V$. For an exactly
periodic distribution the unit vectors would all be identical
so that $V=1$ and $R=0$. For a distribution with no long-range phase
coherence the directions of the unit vectors would be random and so in
the large sample limit $V\sim 0$ and $R\sim 1$. Small values of $R$ are
thus expected for near-periodic distributions.

For the BEKS data, we find $R=0.33$ for a period of 130
$h^{-1}$Mpc obtained as described above. 
Lower values of $R$ are found for 46 samples in {\bf t1},
for 66 samples in {\bf L1}, and for 56 samples in {\bf L2}.
Thus, according to this test
the supercluster distribution in the BEKS data is more periodic
than the CDM models at the 2.4 per cent significance level for 
{\bf t1}, the 2.5 per cent level for {\bf L1} and the 2.1 per cent 
level for {\bf L2}.
The period distribution of the samples with $R\le R$(BEKS) is shown in
Figure \ref{dist_R}. Many of these low-$R$ samples have periods in the range
[100$h^{-1}$Mpc, 140$h^{-1}$Mpc]. Thus in CDM model universes it
is common for supercluster spikes to have a typical separation
similar to that seen in the BEKS data and in a few per cent of cases
the spikes are just as regularly spaced as in the real data.

\section{Discussion and conclusion}
By creating a number of mock pencil-beam surveys we have compared
the apparent periodicity in two CDM model universes with that observed
in the data of Broadhurst et al. (1990).
 The power spectrum analysis alone shows that the BEKS data are 
significantly more periodic than the models at 
about the half per cent level, 
while the PG $\Delta$-test shows less significance, about 10 per
cent for {\bf t1} and 5 per cent for {\bf L1} and {\bf L2}. The supercluster
statistic gives a two per cent probability of finding a structure as
regular as the BEKS data. 
Restricting to a length scale $\sim$100-150$h^{-1}$Mpc,
however, the number of samples which show the kind of periodicity 
seen in the BEKS data is extremely small for each of these statistics. 
Overall no sample is more regular than the BEKS data
for {\it all} three statistics for a single period. 
The two popular CDM models we have studied 
here are apparently unsuccessful in reproducing the observed periodicity.
From this result together with the fact that the statistical results 
appeared to be insensitive to the choice of the bias model, 
we conclude that CDM models conflict 
with the BEKS observation. Either the models need additional physics,
or the data are a fluke or are somehow biased.
 
 Various possible physical explanations have been proposed, 
such as coherent peculiar 
velocities (Hill, Steinhardt and Turner 1991) oscillations in the 
Hubble parameter (Morikawa 1991) or baryonic features in the power
spectrum (Eisenstein \etal 1998) but all of them seem to require either
additional mechanisms with fine tunings beyond the standard theory or 
cosmological parameters significantly different from currently favoured values.
Intriguingly, Dekel et al. (1992) demonstrated that 
built-in power on a large ($\sim 100 h^{-1}$Mpc) length scale 
in the initial density fluctuation could indeed reproduce
periodic features on a given scale, at least by some of the
tests we have considered. If such excess power
on large scales (hence still in the linear regime) exists, 
it will be detectable in the power spectra of the future
2dF and Sloan surveys.

 Having found at least a few examples that are nearly 
as periodic as the BEKS data, we cannot rule out the possibility
that the BEKS data (or the Galactic Pole direction) are a fluke. 
On the other hand, one should be aware of the 
complexity of the original observations --an incomplete compilation of
a narrow and deep, and of a wide and shallow survey at each of the
Galactic Poles. It is not clear whether such a combination constitutes a fair sample. 
No evidence for periodic
structure on $\sim$130$h^{-1}$Mpc has been found so far in two other
deep redshift surveys, the 
ESO-Sculptor Survey (Bellanger and de Lapparent 1995) and 
the Caltech Faint Galaxy Redshift Survey (Cohen 1999).
Follow-up observations to BEKS by Koo \etal (1993) did {\it not} show
a strong regularity in two other directions, although around 
the Galactic Pole the regularity was found to be further strengthened.
Our results give the {\it a priori} probability for such apparent
periodicity in CDM models. Several more deep surveys might
suffice to judge whether the discrepancy with BEKS reflects a major 
inconsistency.
The planned VIRMOS Deep Survey (Le F\`{e}vre \etal 1998, see also Guzzo
1999) will survey the range $0.3\le z\le 1$ and will
provide, together with the large volume 2dF and Sloan surveys,
much larger and more complete samples in the near future.

\section*{Acknowledgments}
We are grateful to  Carlton Baugh and Shaun Cole for fruitful discussions. 
We thank the referee, Avishai Dekel, for giving us constructive comments
on the manuscript. 
The simulations were carried out on the Cray-T3E at the Rechenzentrum,
Garching (RZG).

\bsp

\label{lastpage}

\clearpage

\onecolumn

\clearpage

\vspace*{-1cm}

\begin{table}
\begin{center}
\begin{tabular}{c}
Figure 5: The BEKS data\\
\\
{\large $\max\{P(k)\}$=0.177 at $k=7.5\;\;\;\;\;\;\;$}\\
{\large $\Delta=0.165$ with period $130h^{-1}$Mpc}\\
{\large $R=0.330$ with period $130h^{-1}$Mpc}\\
\epsfig{file=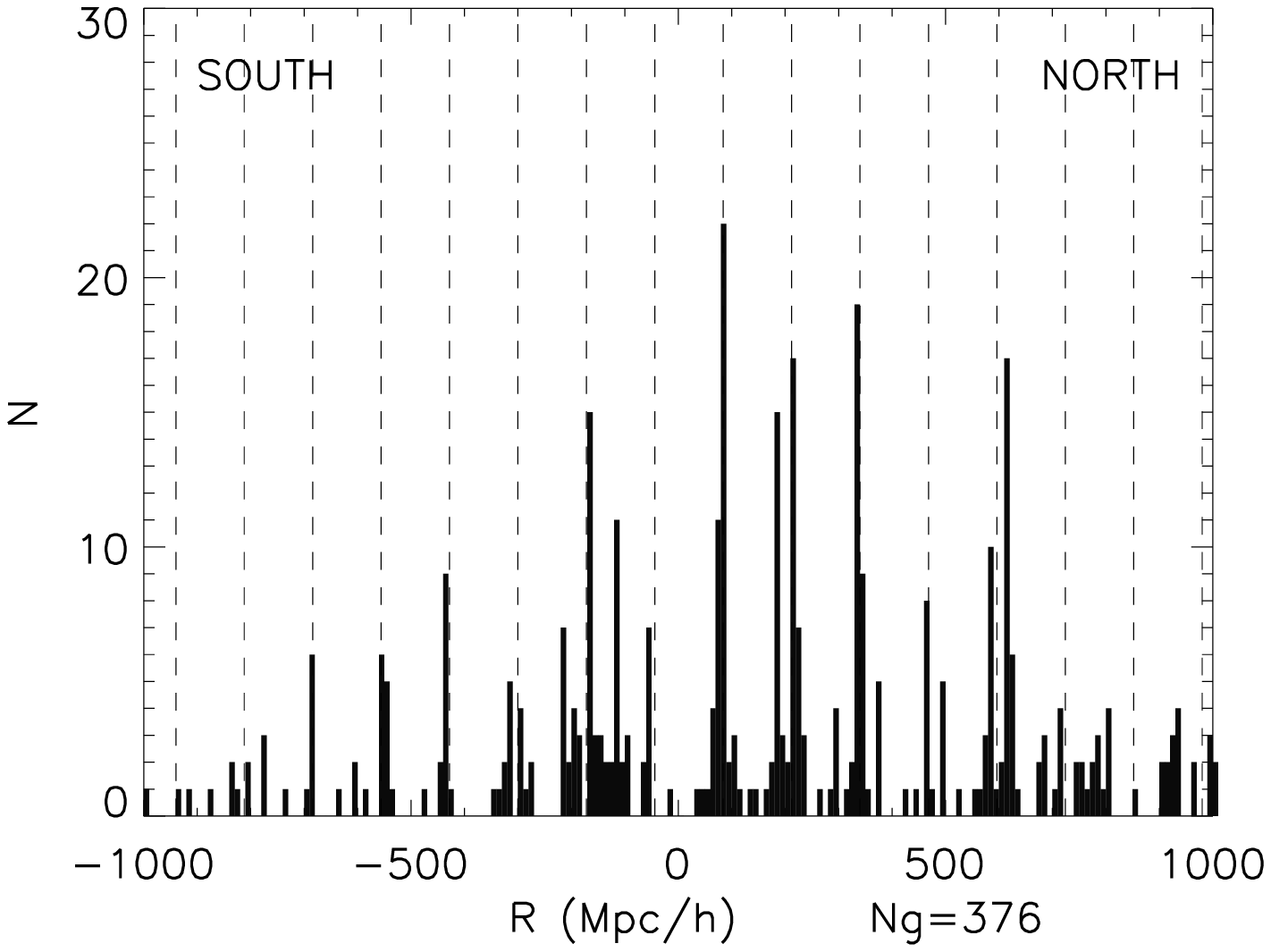,width=12cm,height=5.5cm} \\ 
 \\
\epsfig{file=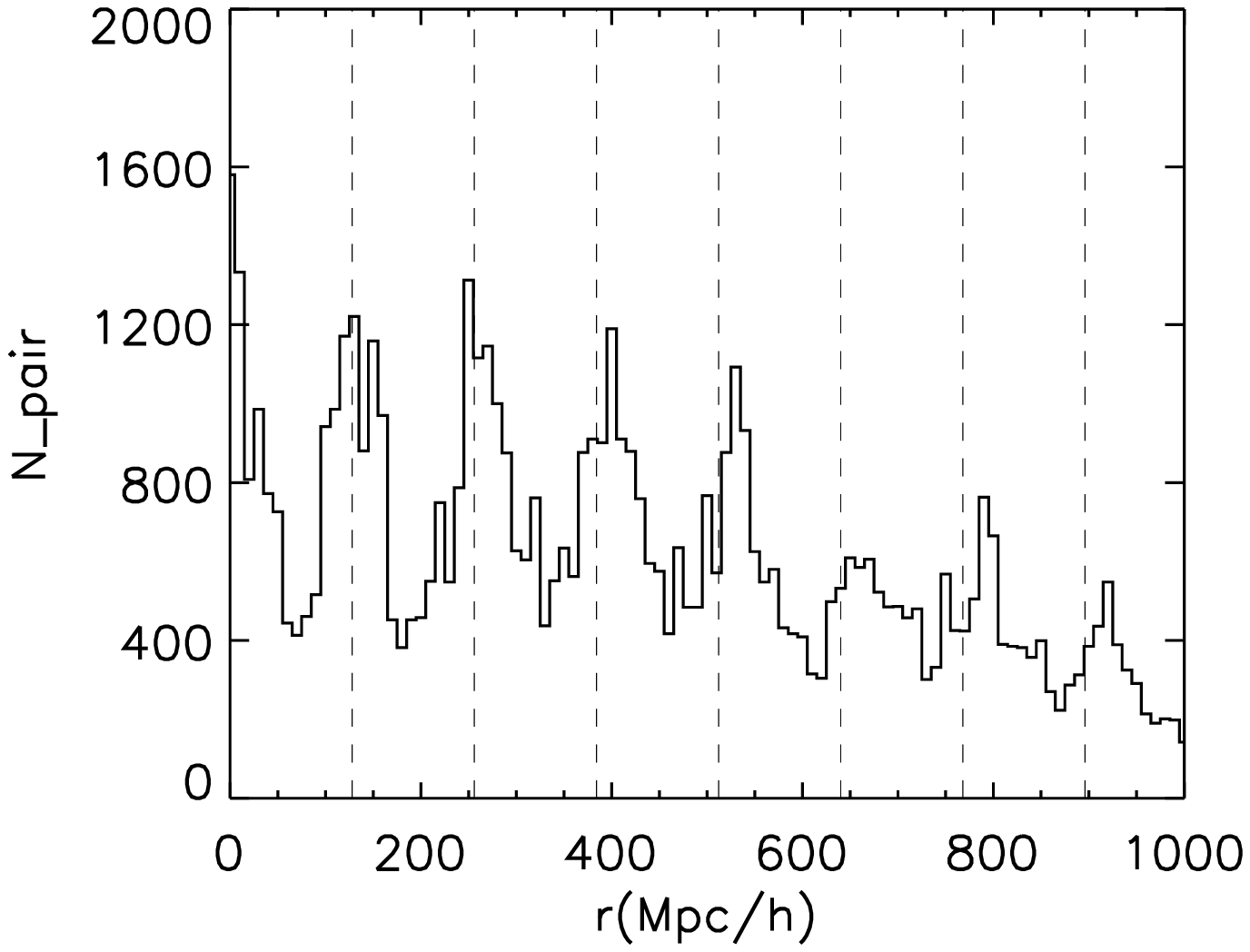,width=12.5cm,height=5.5cm} \\
 \\
\epsfig{file=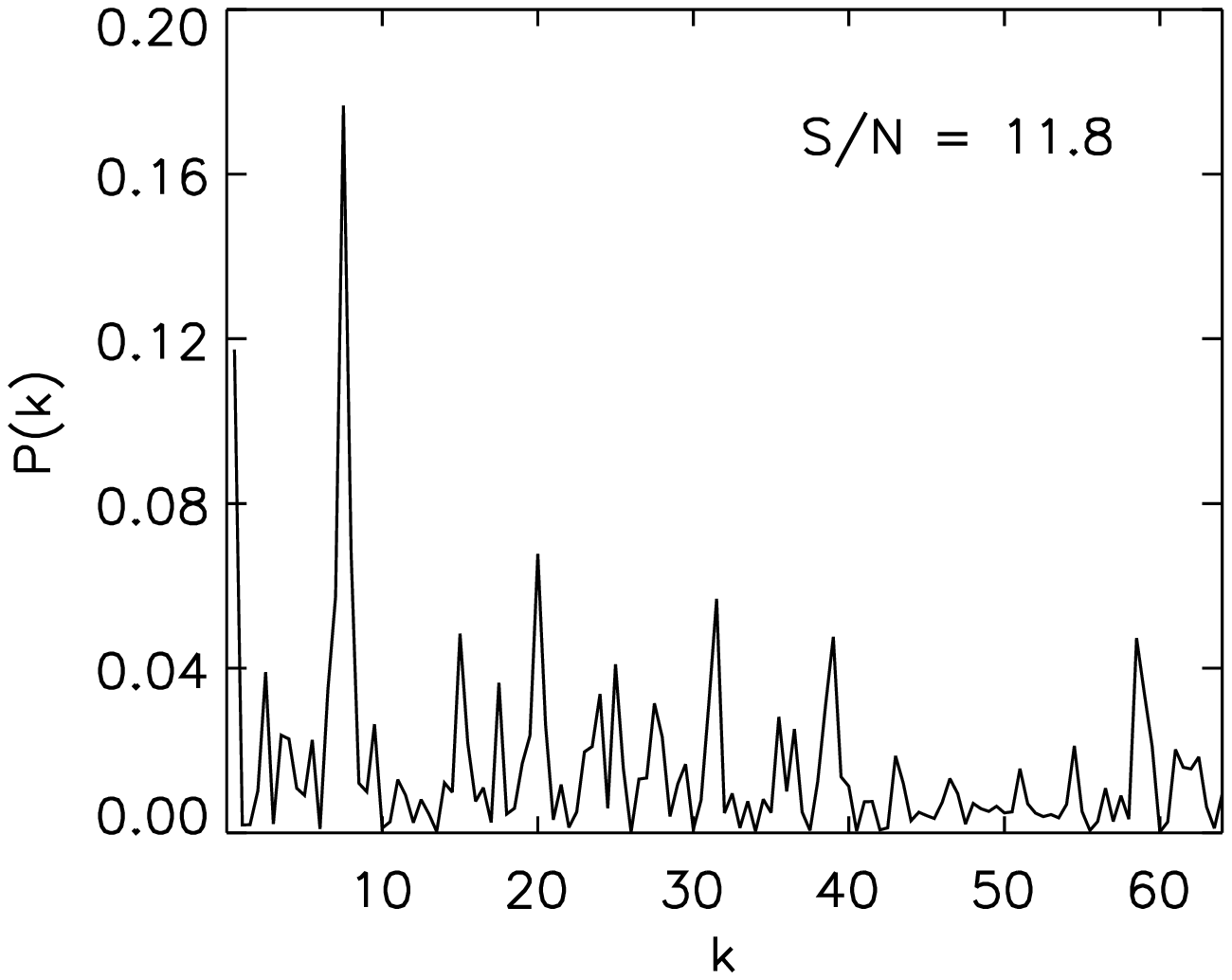,width=12.5cm,height=6cm}
\end{tabular}
\end{center}
\end{table}

\begin{figure}
\label{BEKSdata}
\end{figure}

\begin{figure}
\label{set}
\end{figure}

\begin{table}
\begin{center}
\begin{tabular}{c}
Figure 6: Selected samples. (a) $\tau$CDM model t1.\\
\\
{\large $\max\{P(k)\}$=0.145 at $k=8.5\;\;\;\;\;\;\;$}\\
{\large $\Delta=0.164$ with period $120h^{-1}$Mpc}\\
{\large $R=0.574$ with period $70h^{-1}$Mpc$\;\;$}\\
\epsfig{file=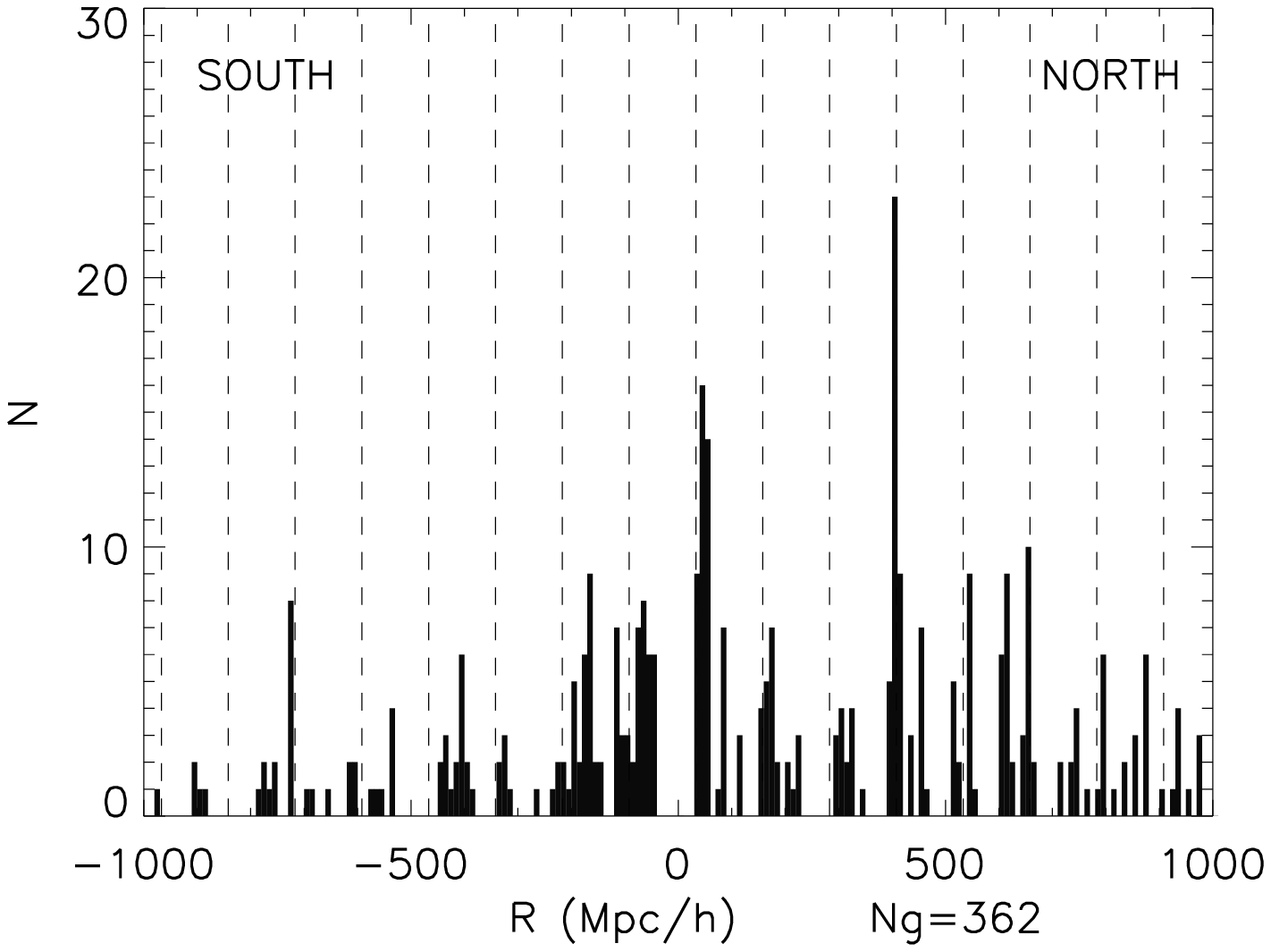,width=12cm,height=5.5cm} \\ 
 \\
\epsfig{file=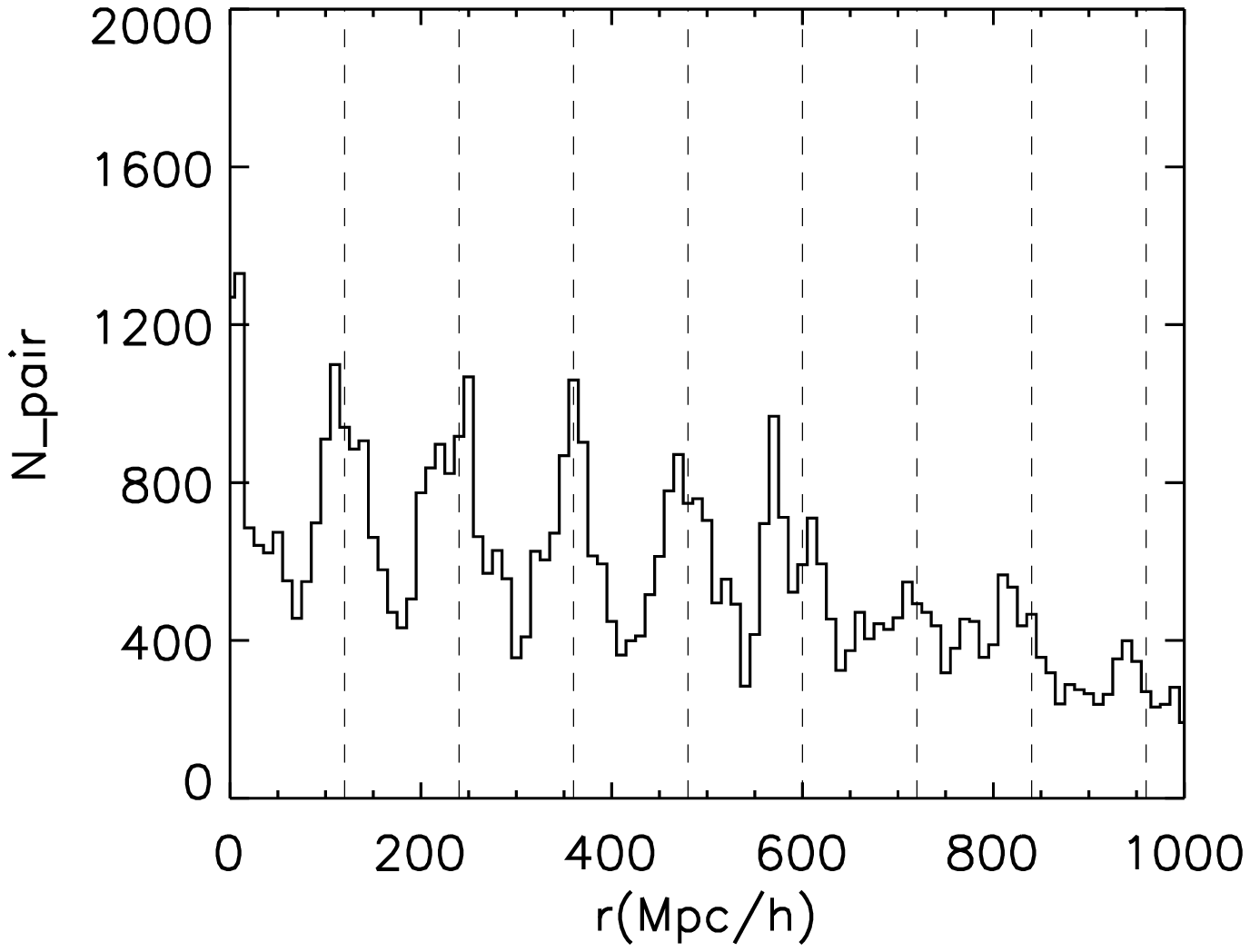,width=12.5cm,height=5.5cm} \\
 \\
\epsfig{file=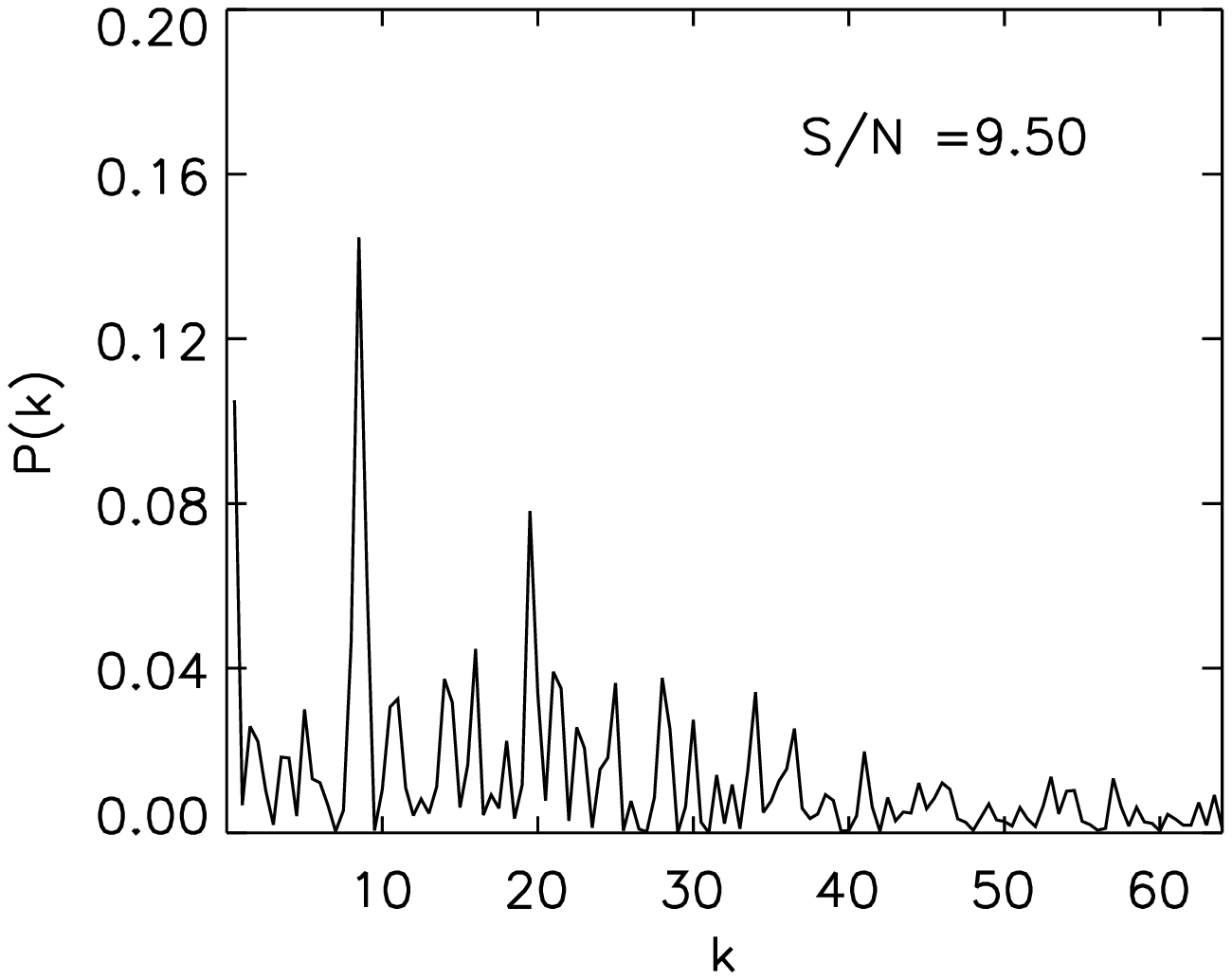,width=12.5cm,height=6cm} \\
\end{tabular}
\end{center}
\end{table}

\begin{table}
\begin{center}
\begin{tabular}{c}
(b) $\tau$CDM model t1.\\
\\
{\large $\max\{P(k)\}$=0.149 at $k=16\;\;\;\;\;\;\;\;$}\\
{\large $\Delta=0.167$ with period $220h^{-1}$Mpc}\\
{\large $R=0.576$ with period $190h^{-1}$Mpc}\\
\epsfig{file=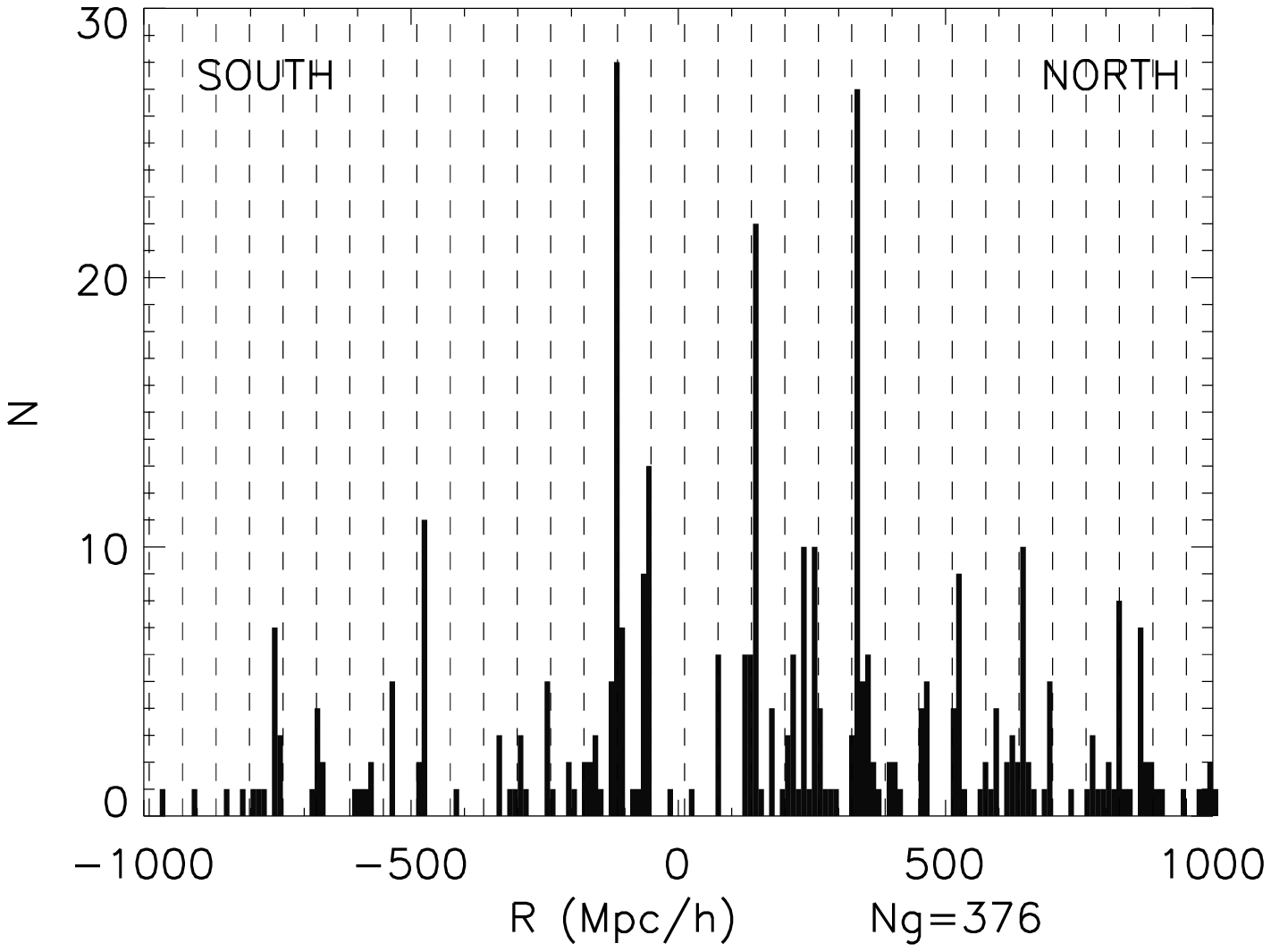,width=12cm,height=5.5cm}\\ 
 \\
\epsfig{file=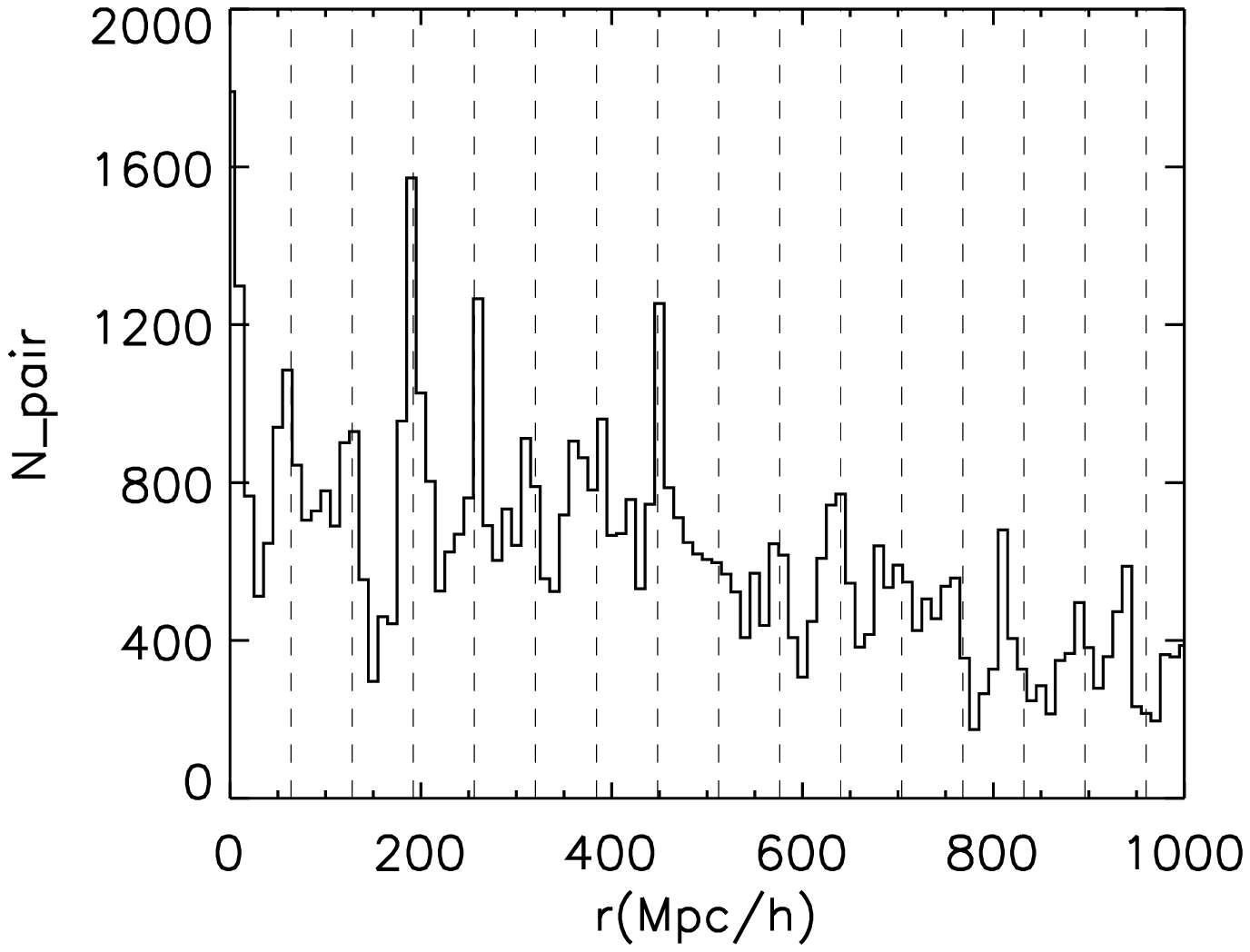,width=12.5cm,height=5.5cm}\\ 
 \\
\epsfig{file=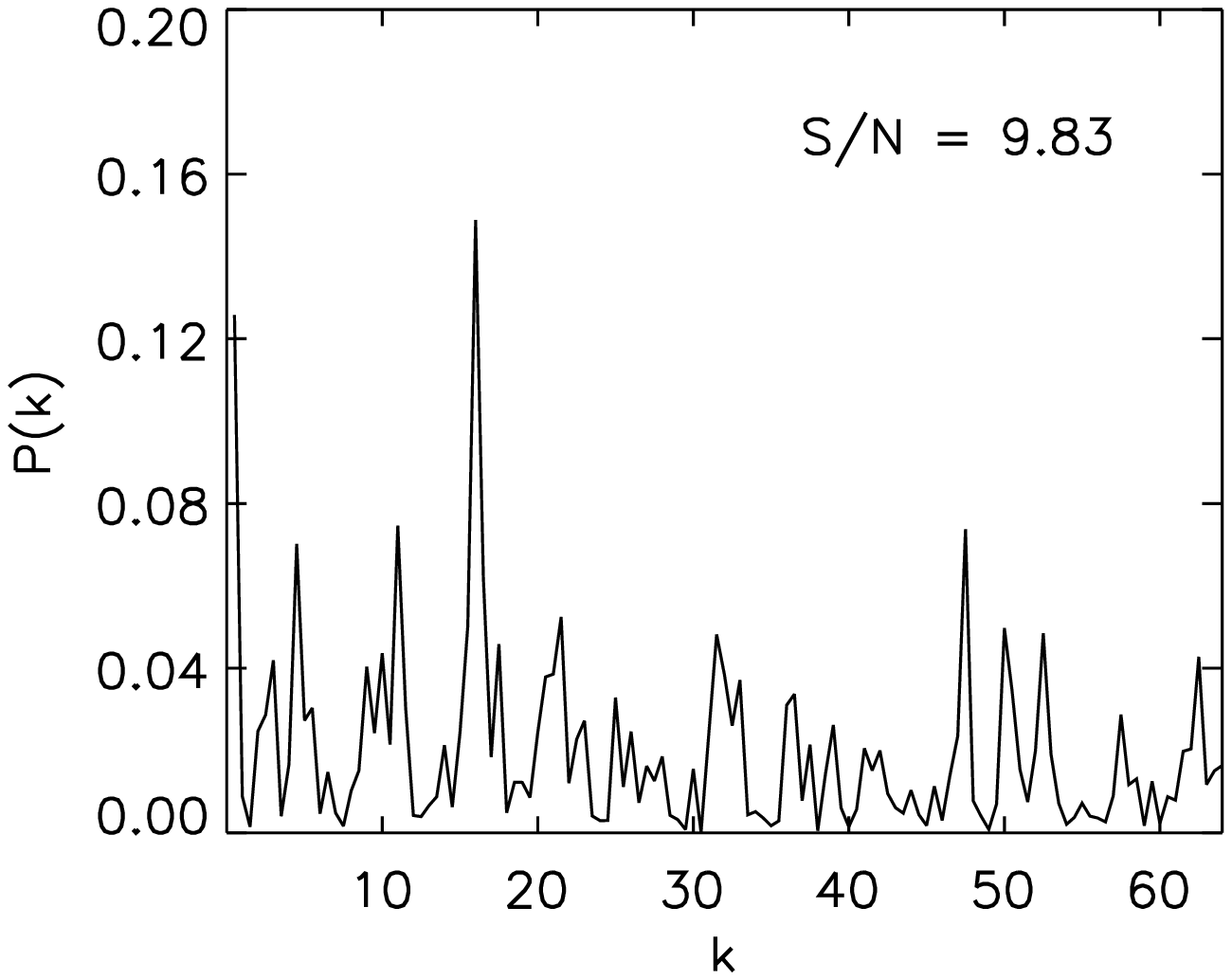,width=12.5cm,height=6cm}\\ 
\end{tabular}
\end{center}
\end{table}

\begin{table}
\begin{center}
\begin{tabular}{c}
(c) $\tau$CDM model t1\\
\\
{\large $\max\{P(k)\}$=0.201 at $k=2.5\;\;\;\;\;\;\;$}\\
{\large $\Delta=0.152$ with period $830h^{-1}$Mpc}\\
{\large $R=0.505$ with period $160h^{-1}$Mpc}\\
\epsfig{file=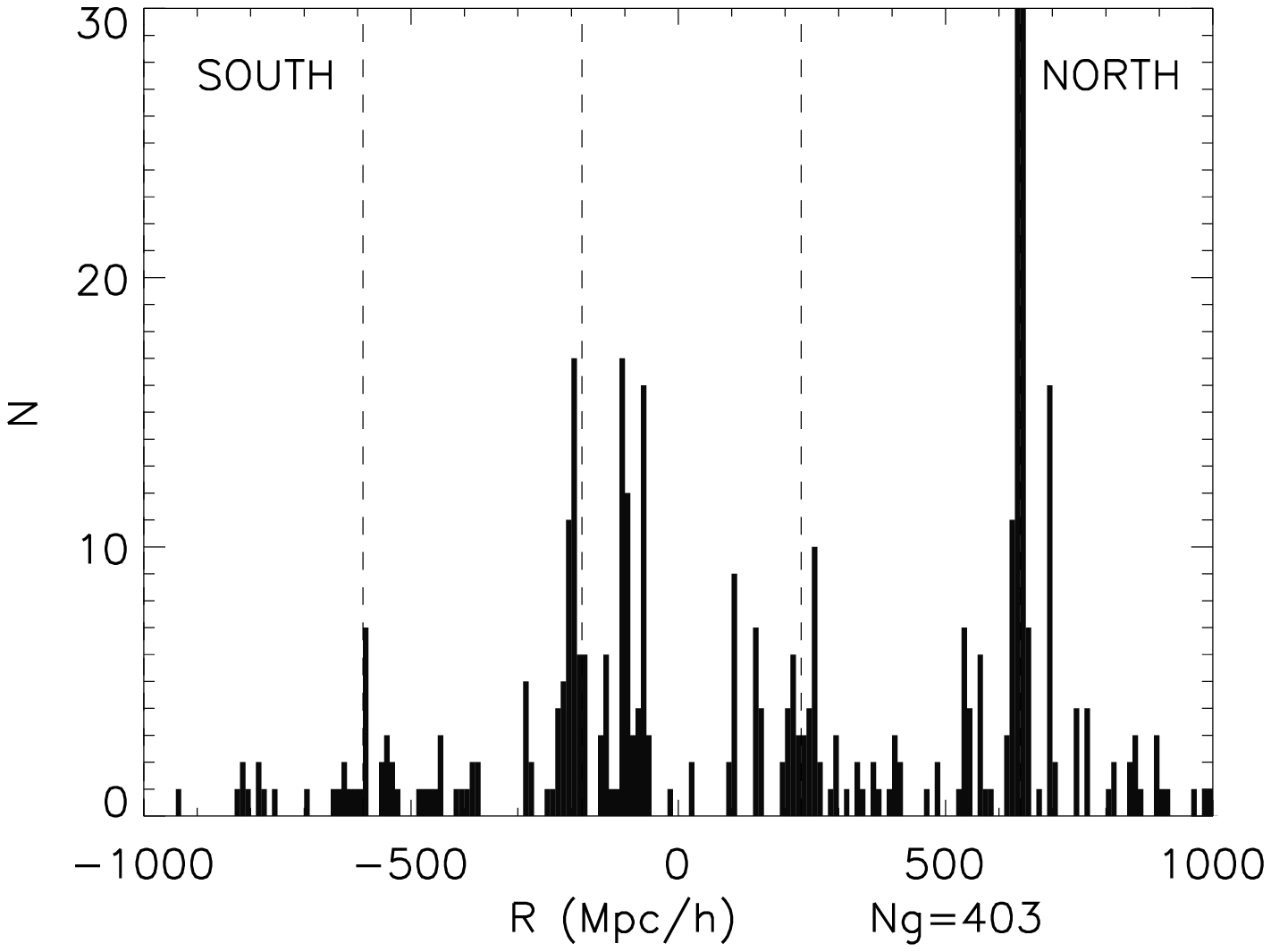,width=12cm,height=5.5cm}\\ 
 \\
\epsfig{file=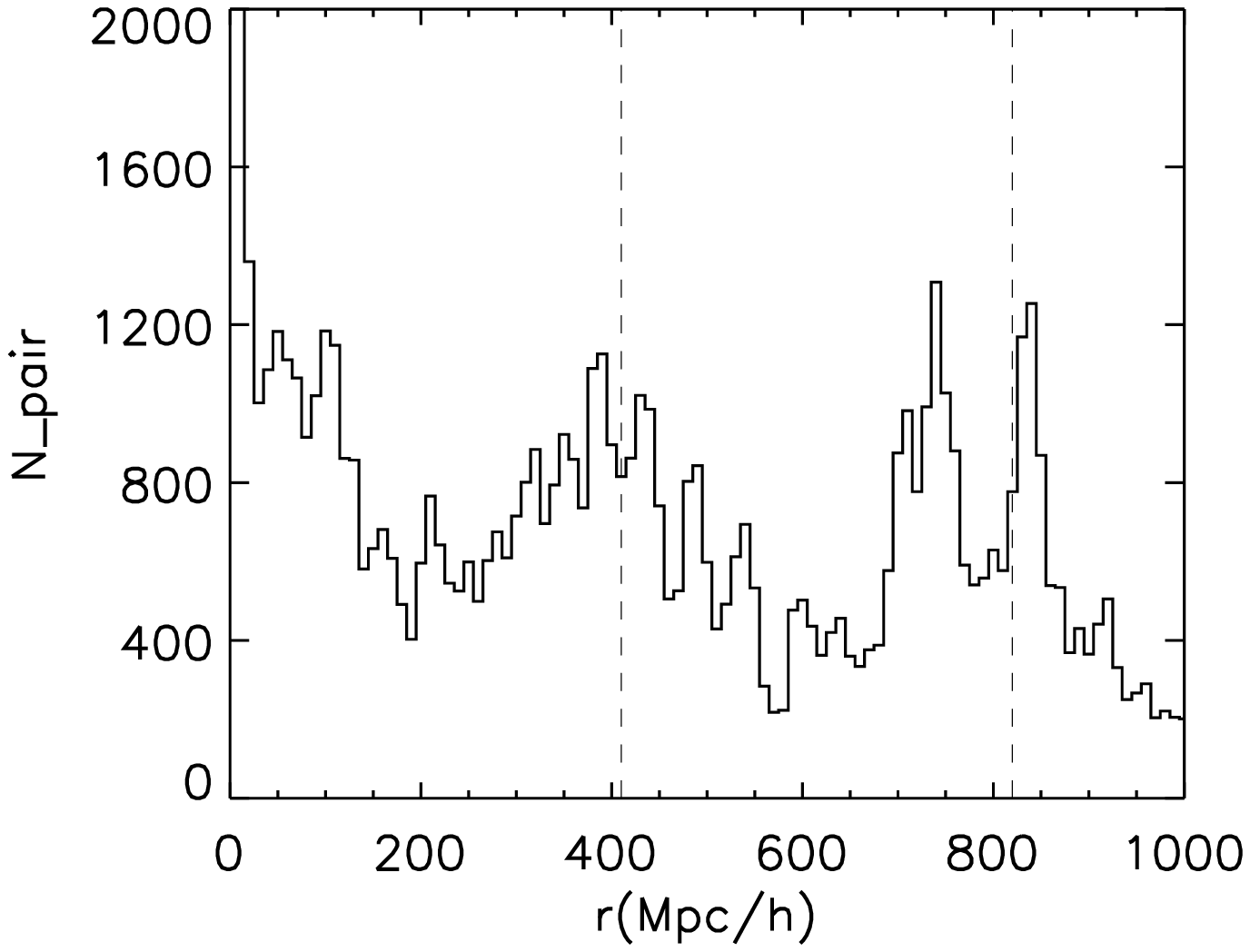,width=12.5cm,height=5.5cm}\\ 
 \\
\epsfig{file=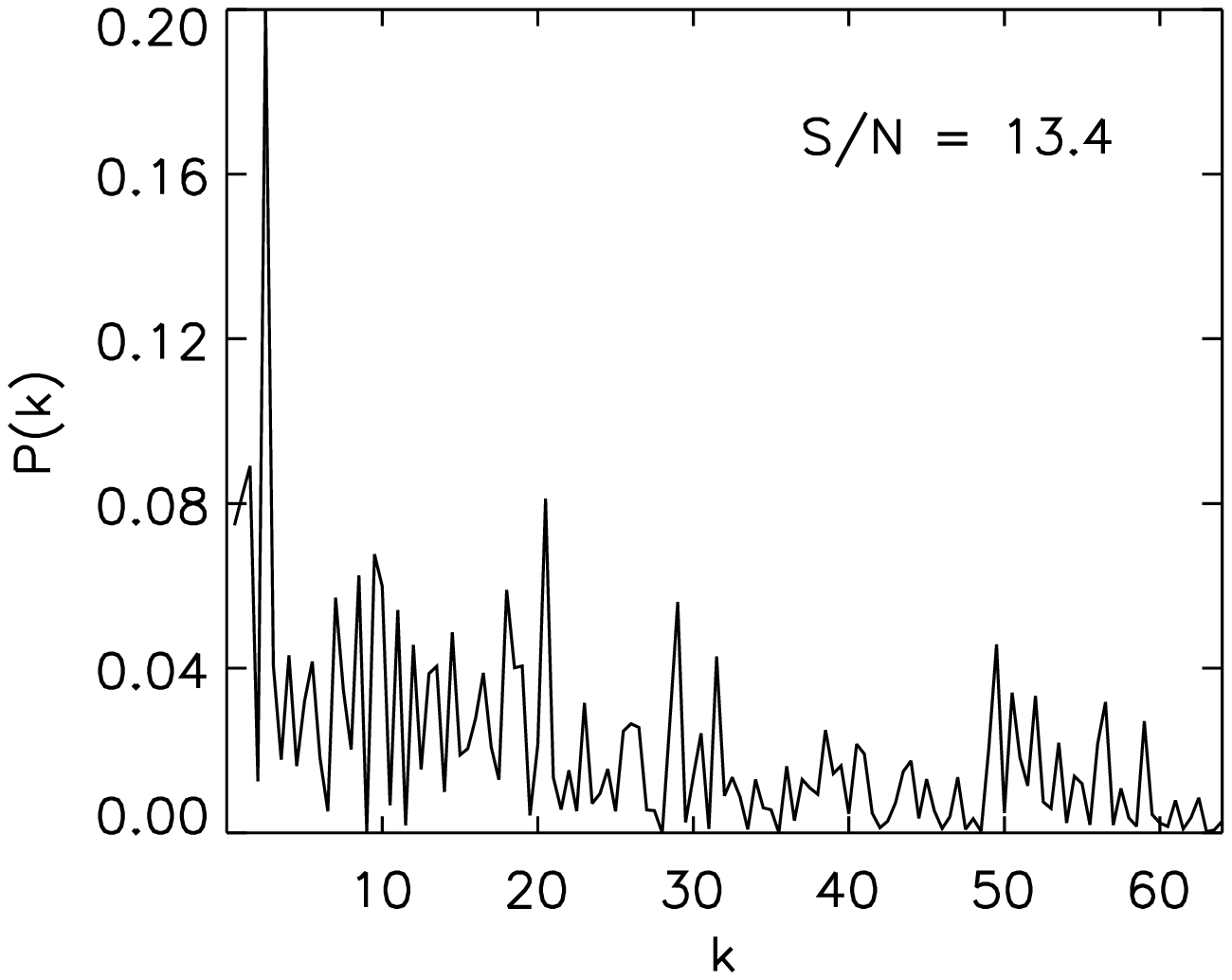,width=12.5cm,height=6cm}\\ 
\end{tabular}
\end{center}
\end{table}

\begin{table}
\begin{center}
\begin{tabular}{c}
(d) $\Lambda$CDM model L1\\
\\
{\large $\max\{P(k)\}$=0.150 at $k=5.0\;\;\;\;\;\;\;$}\\
{\large $\Delta=0.159$ with period $200h^{-1}$Mpc}\\
{\large $R=0.358$ with period $200h^{-1}$Mpc}\\
\epsfig{file=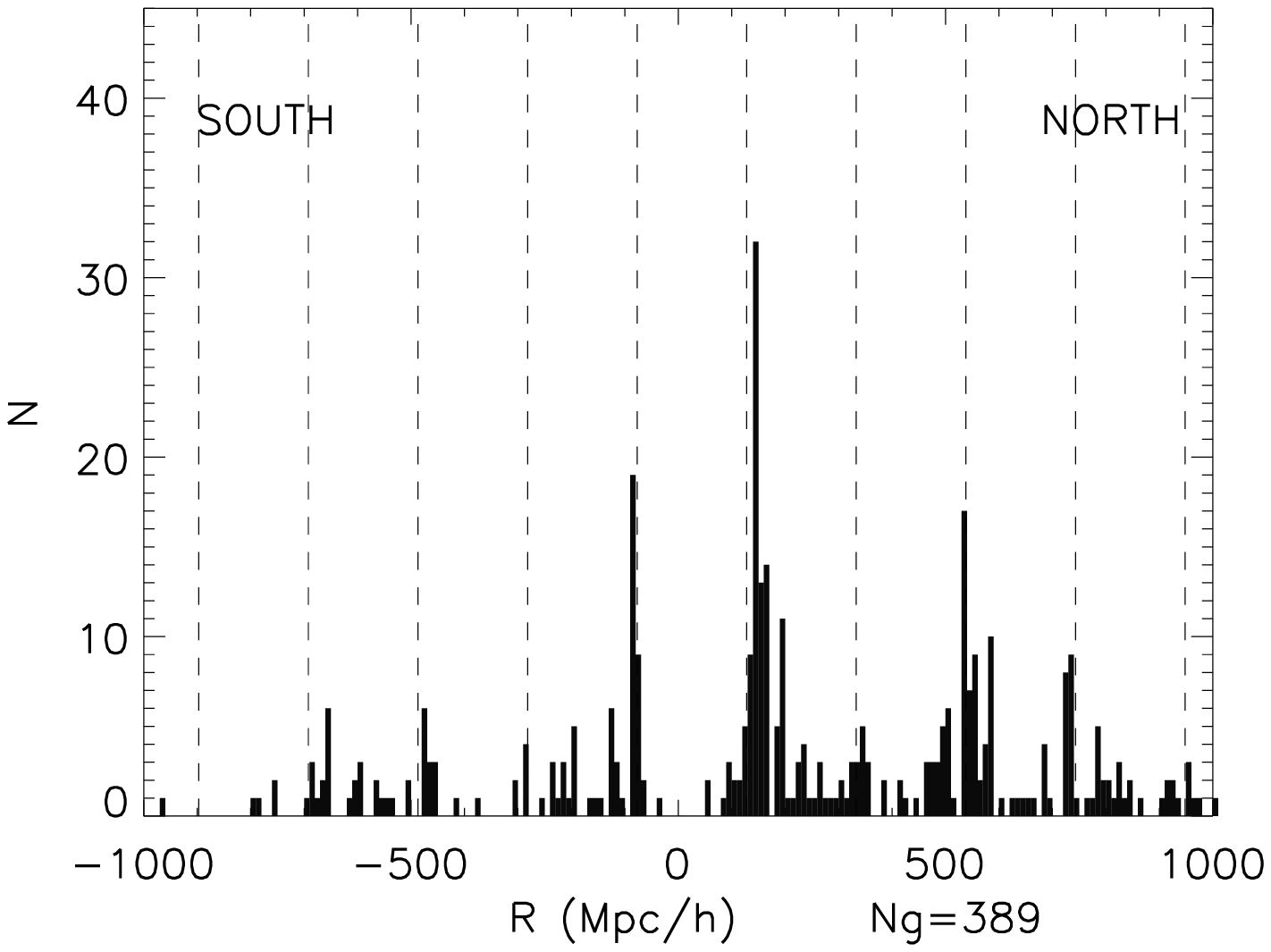,width=12cm,height=5.5cm}\\ 
 \\
\epsfig{file=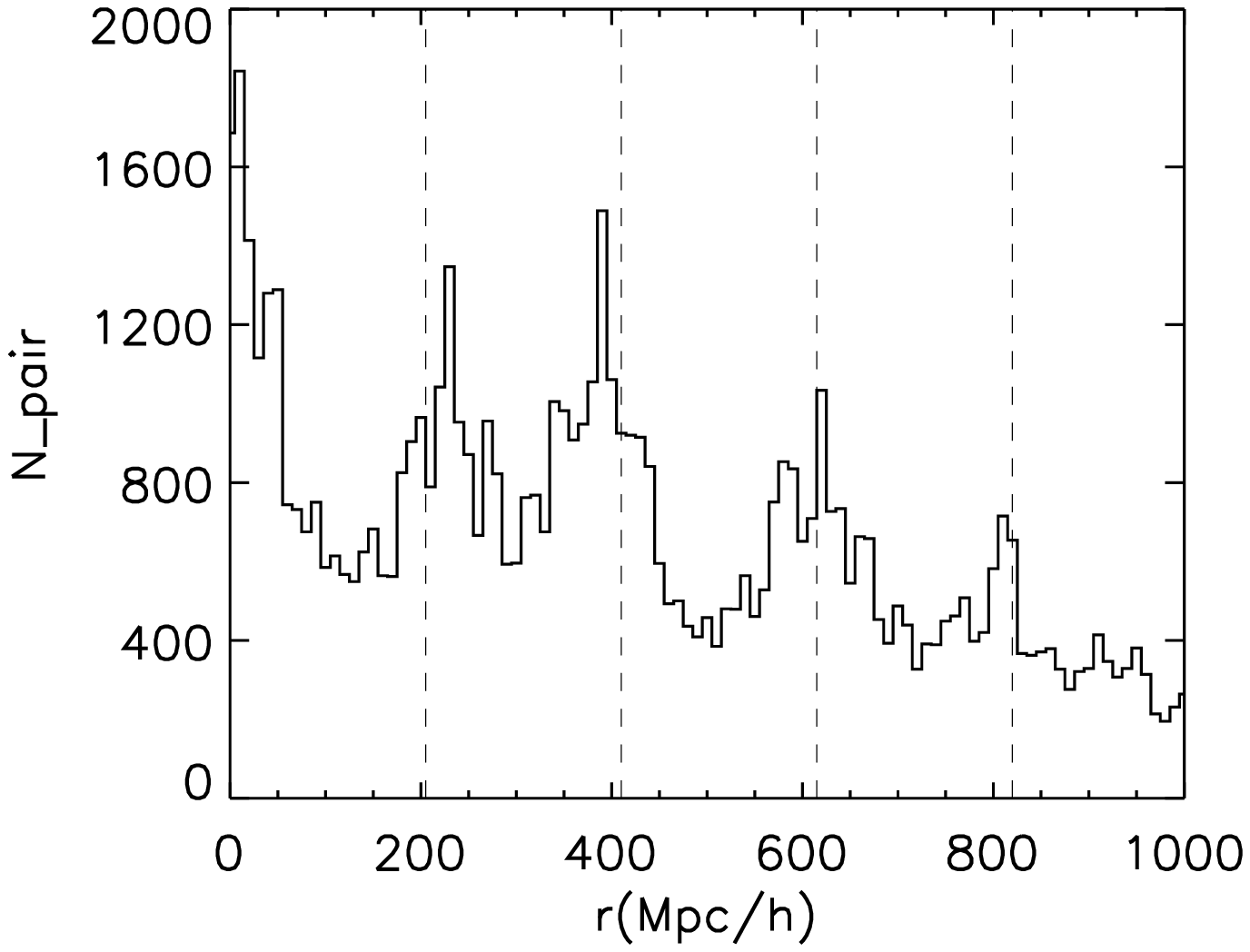,width=12.5cm,height=5.5cm}\\ 
 \\
\epsfig{file=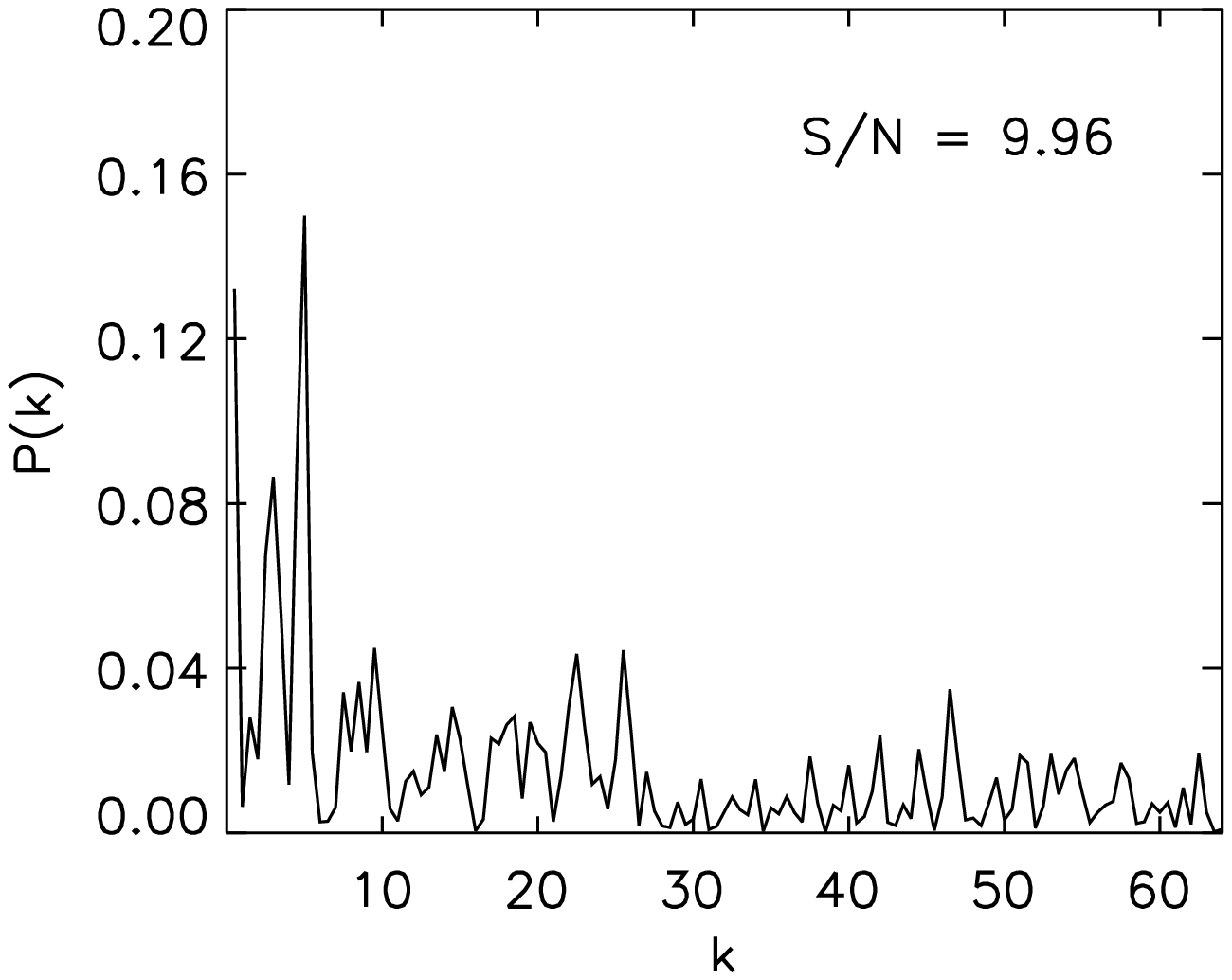,width=12.5cm,height=6cm}\\ 
\end{tabular}
\end{center}
\end{table}

\begin{table}
\begin{center}
\begin{tabular}{c}
(e) $\Lambda$CDM model L2\\
\\
{\large $\max\{P(k)\}$=0.186 at $k=10.5\;\;\;\;$}\\
{\large $\Delta=0.160$ with period $90h^{-1}$Mpc}\\
{\large $R=0.475$ with period $90h^{-1}$Mpc}\\
\epsfig{file=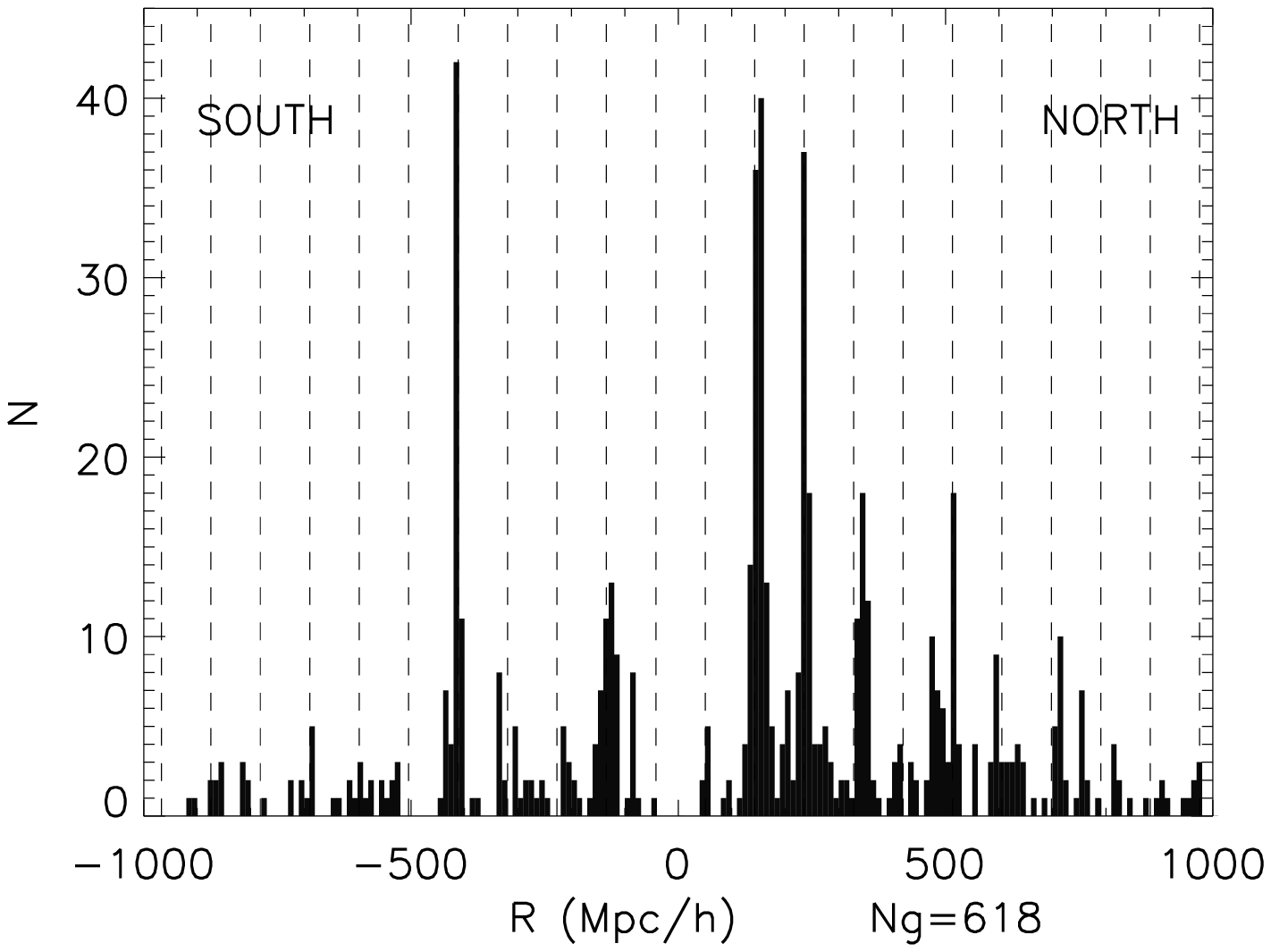,width=12cm,height=5.5cm}\\ 
 \\
\epsfig{file=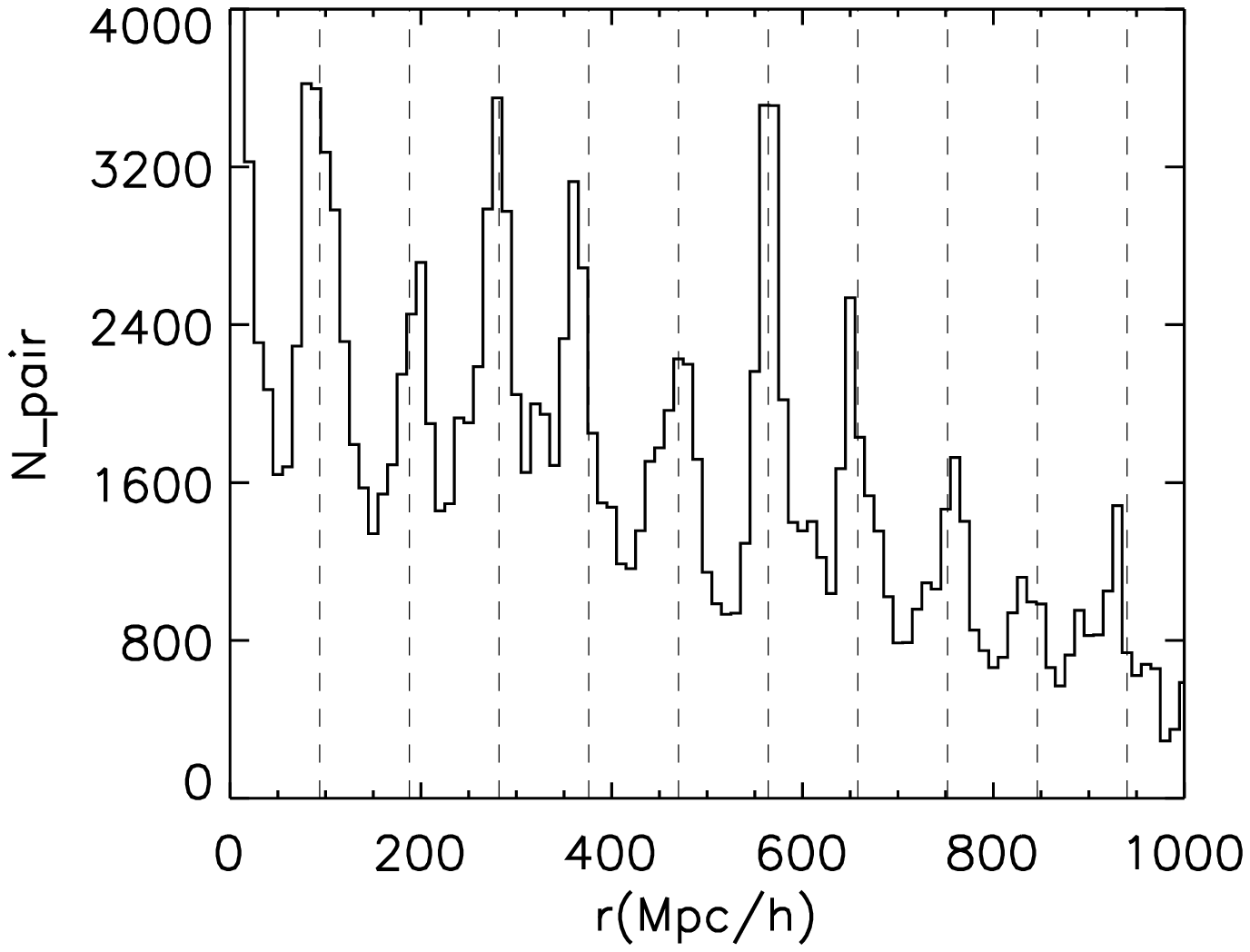,width=12.5cm,height=5.5cm}\\ 
 \\
\epsfig{file=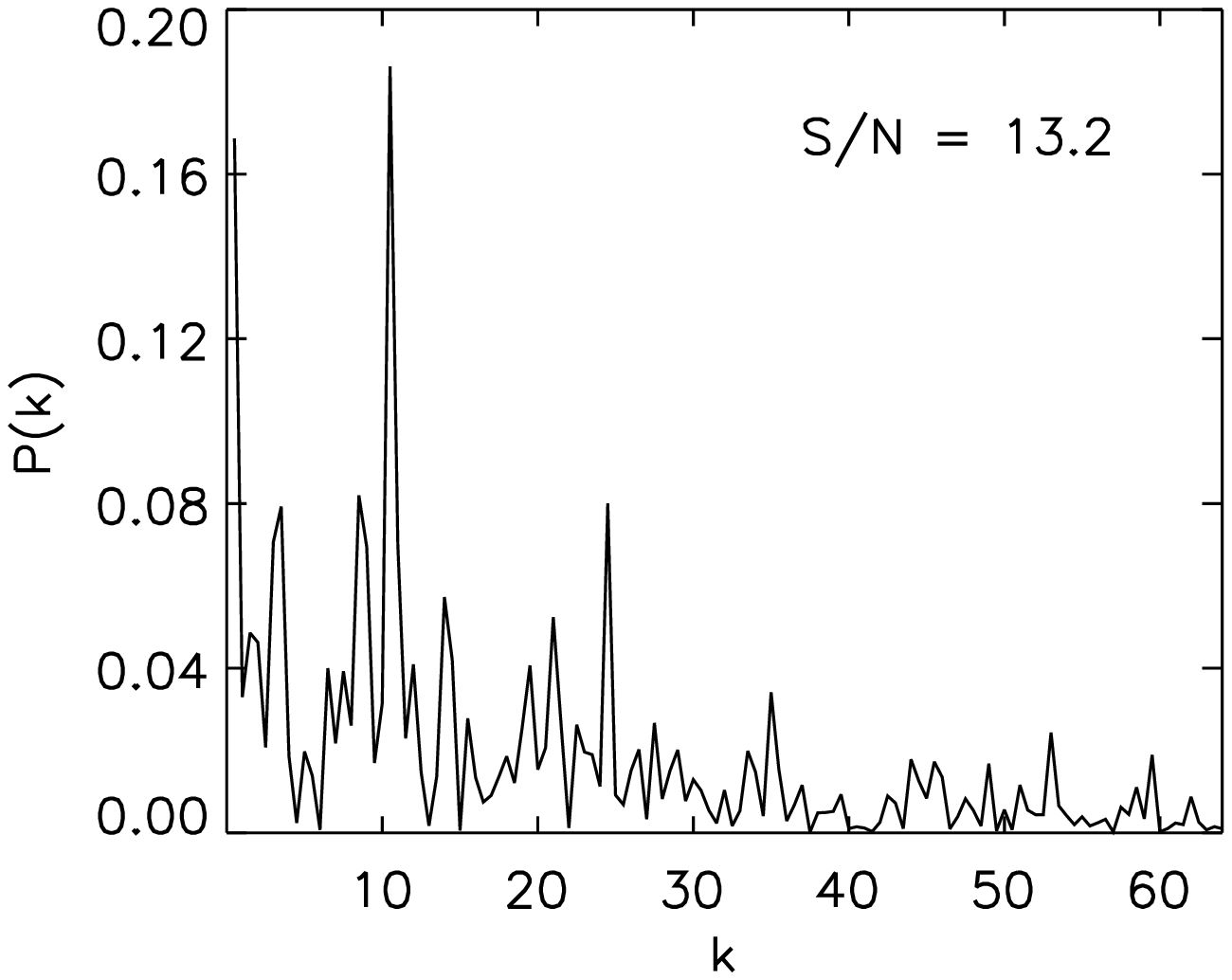,width=12.5cm,height=6cm}\\ 
\end{tabular}
\end{center}
\end{table}

\begin{table}
\begin{center}
\begin{tabular}{c}
(f) $\Lambda$CDM model L2\\
\\
{\large $\max\{P(k)\}$=0.176 at $k=5.0\;\;\;\;\;\;\;$}\\
{\large $\Delta=0.155$ with period $190h^{-1}$Mpc}\\
{\large $R=0.572$ with period $190h^{-1}$Mpc}\\
\epsfig{file=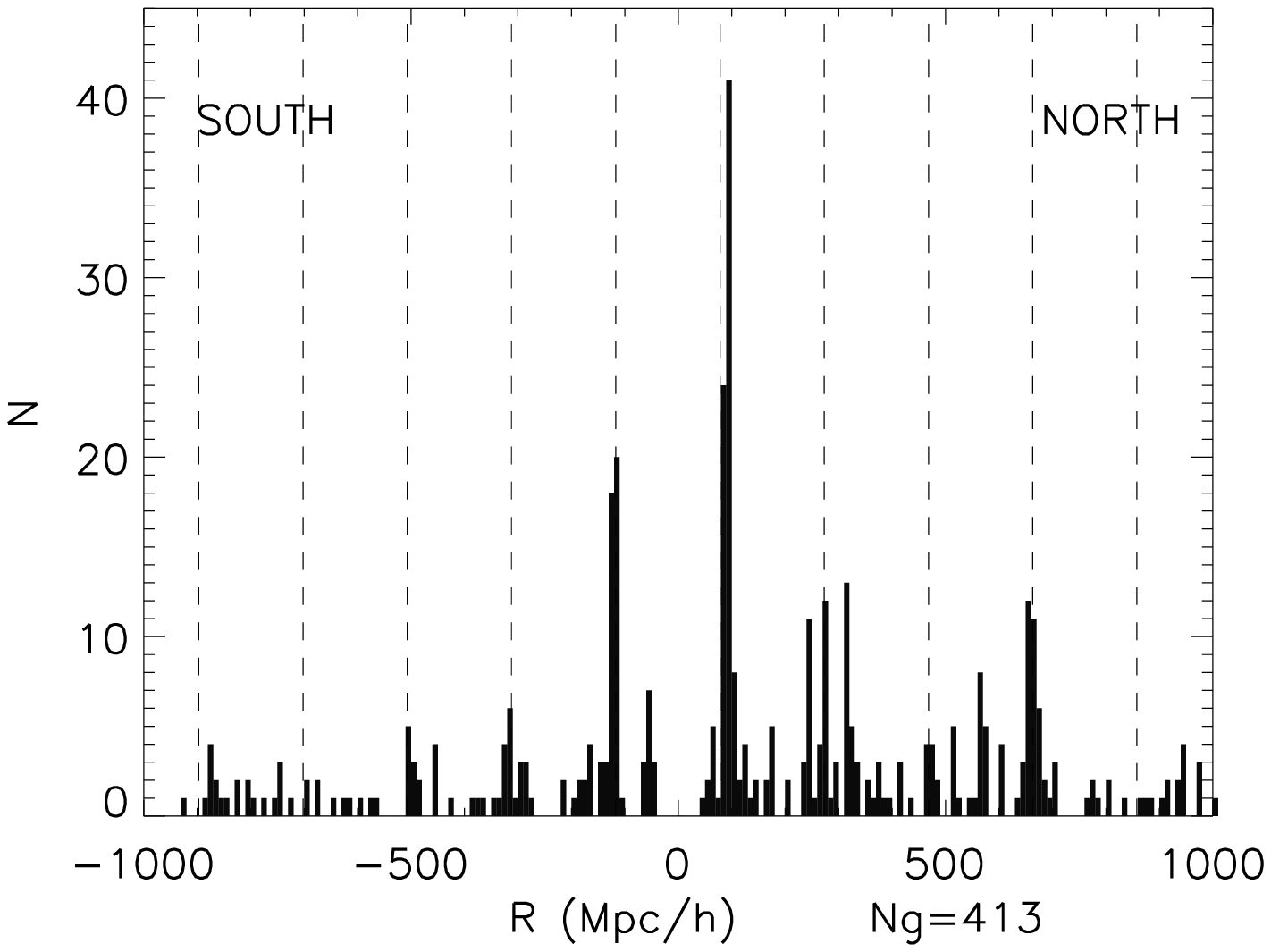,width=12cm,height=5.5cm}\\ 
 \\
\epsfig{file=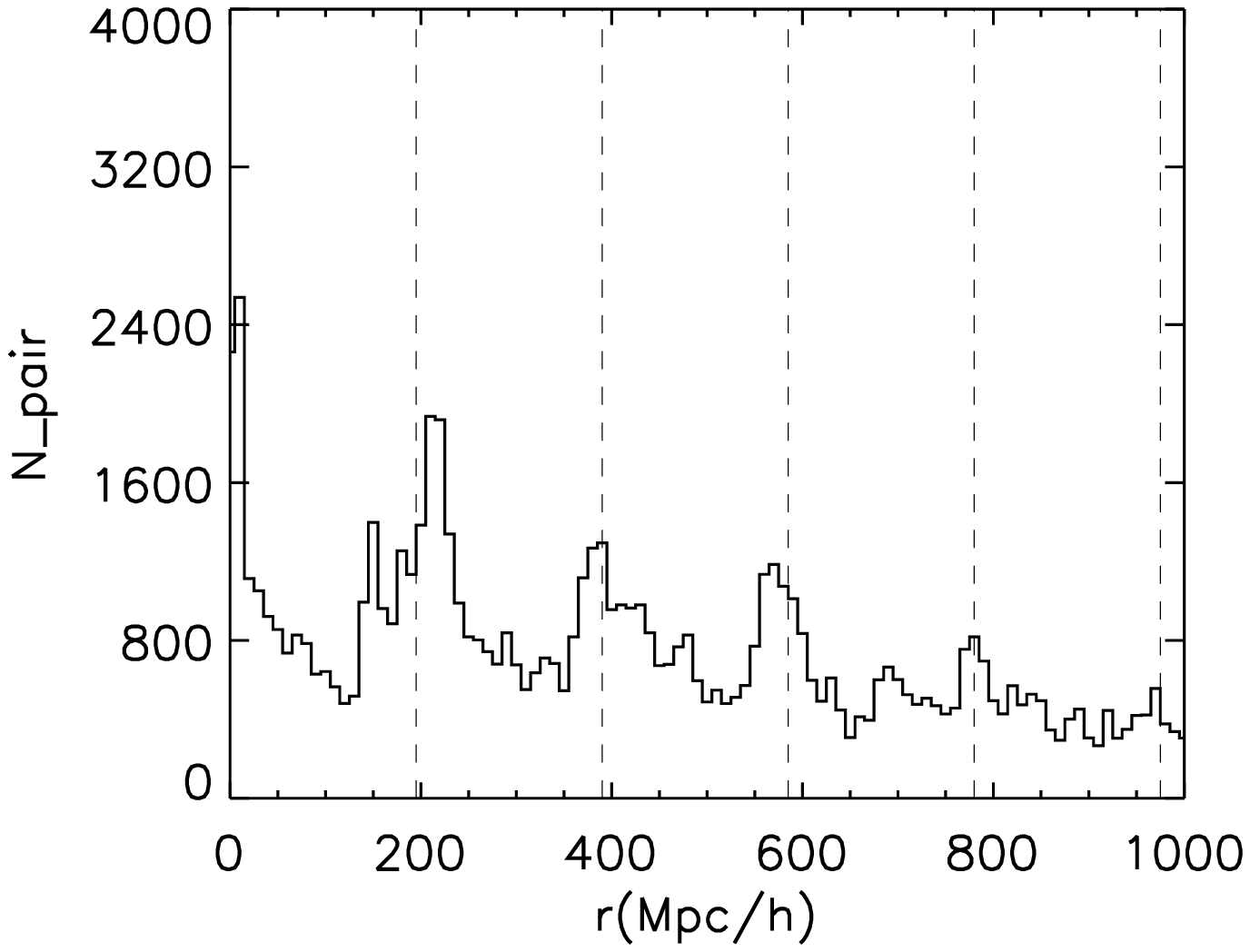,width=12.5cm,height=5.5cm}\\ 
 \\
\epsfig{file=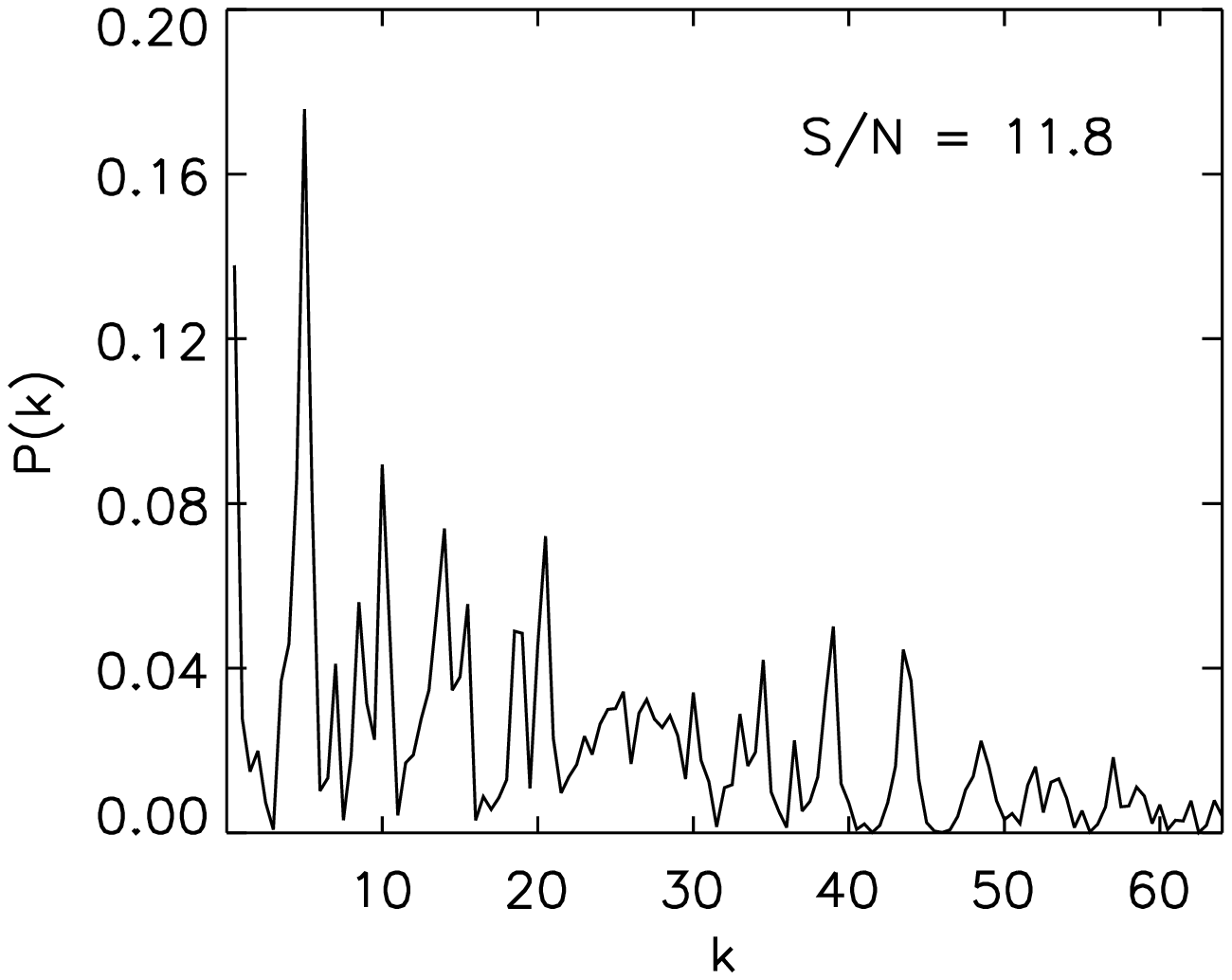,width=12.5cm,height=6cm}\\ 
\end{tabular}
\end{center}
\end{table}

\setcounter{figure}{6}

\begin{figure}
\begin{center}
\epsfig{file=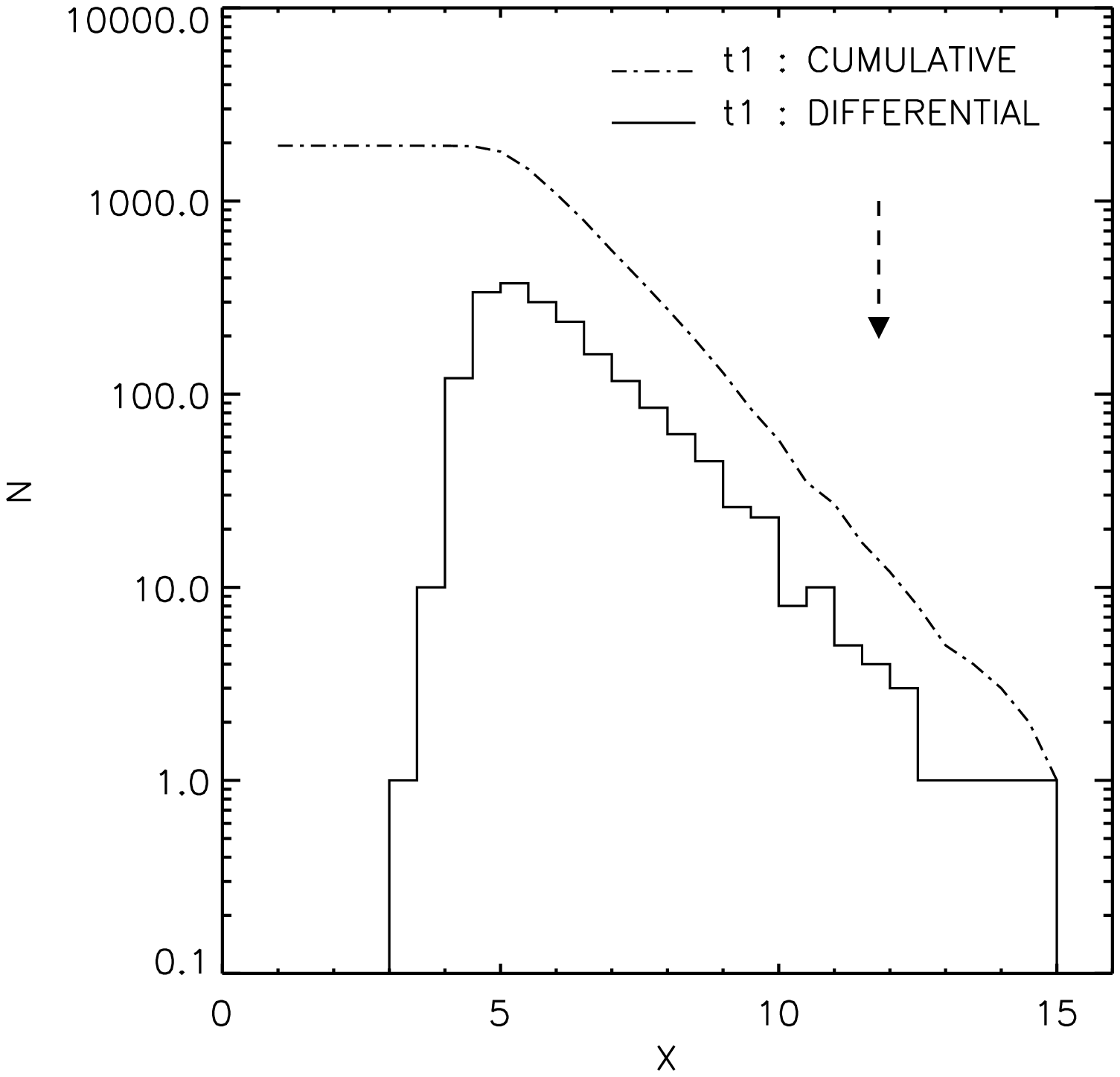,width=8cm,height=6cm}\\
\epsfig{file=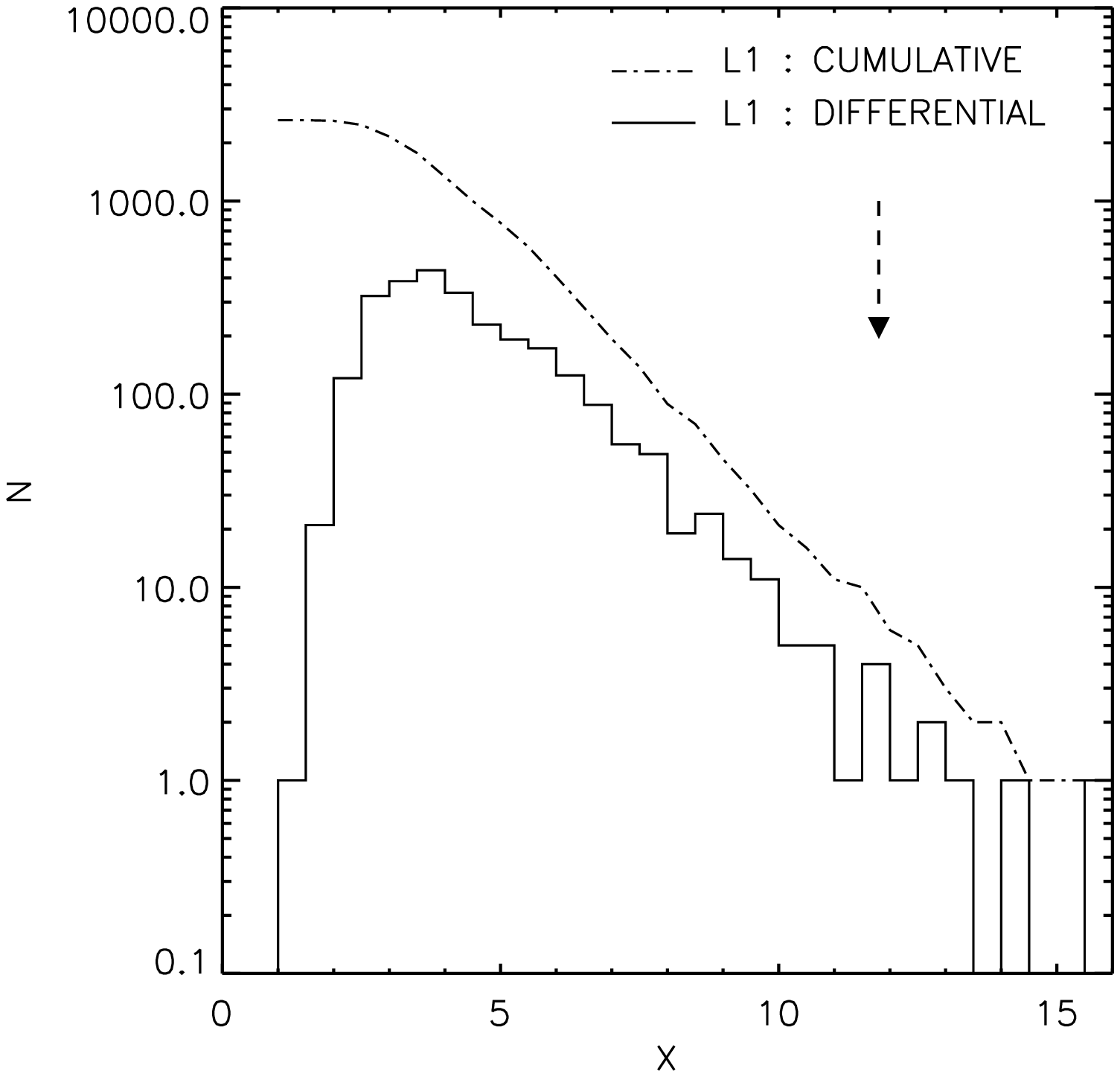,width=8cm,height=6cm}\\ 
\epsfig{file=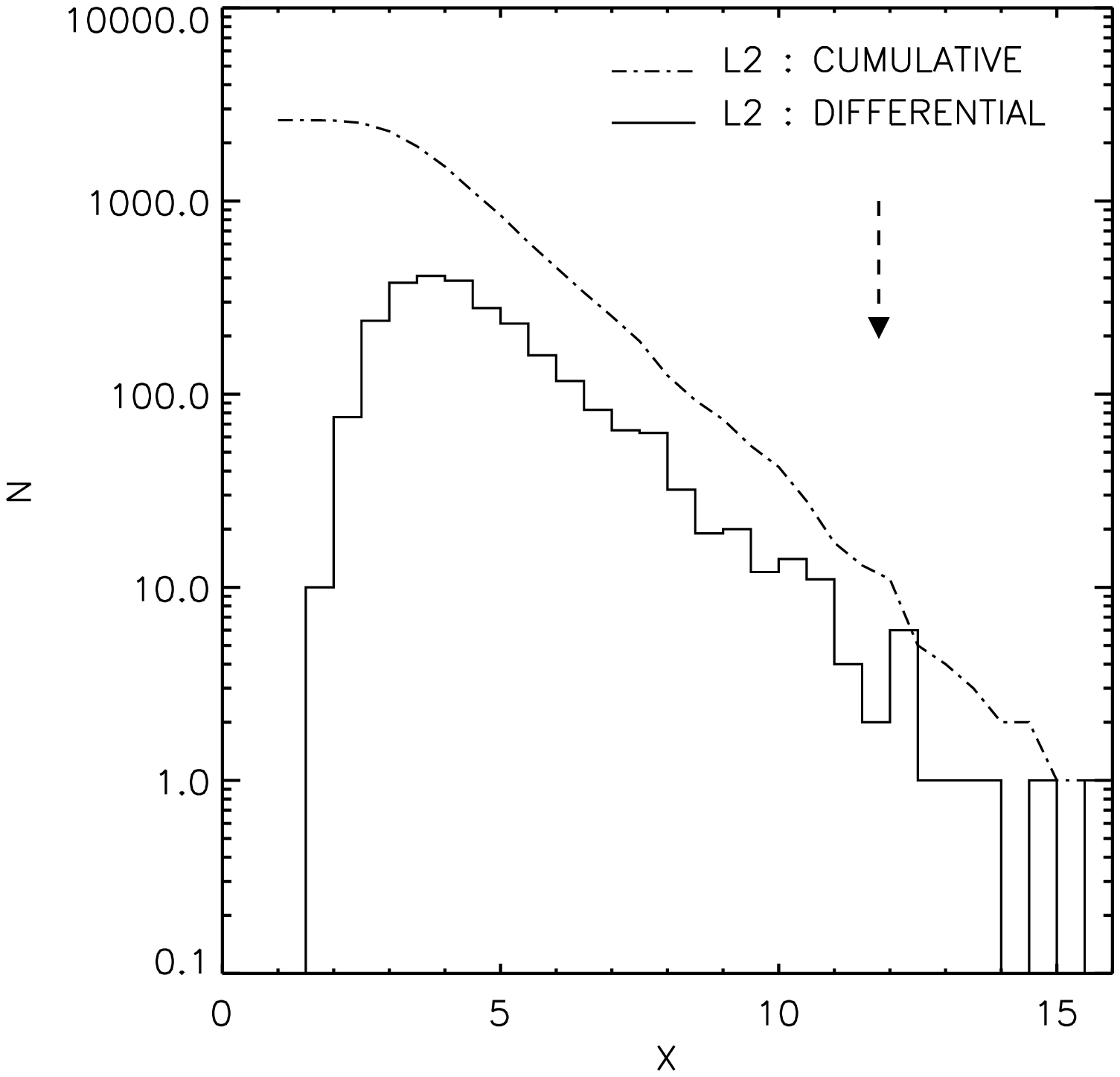,width=8cm,height=6cm} 
\caption{The cumulative(histogram) and the differential(dot-dashed line) 
distribution of the peak-to-noise ratio are plotted for $\tau$CDM bias
{\bf t1}(top), 
$\Lambda$CDM bias {\bf L1}(middle), and bias {\bf L2}(bottom). 
Note the difference in the total number of samples in the two models.
The arrow indicates the peak-to-noise ratio of the BEKS data.}
\label{c_d}
\end{center}
\end{figure}

\begin{figure}
\begin{center}
\epsfig{file=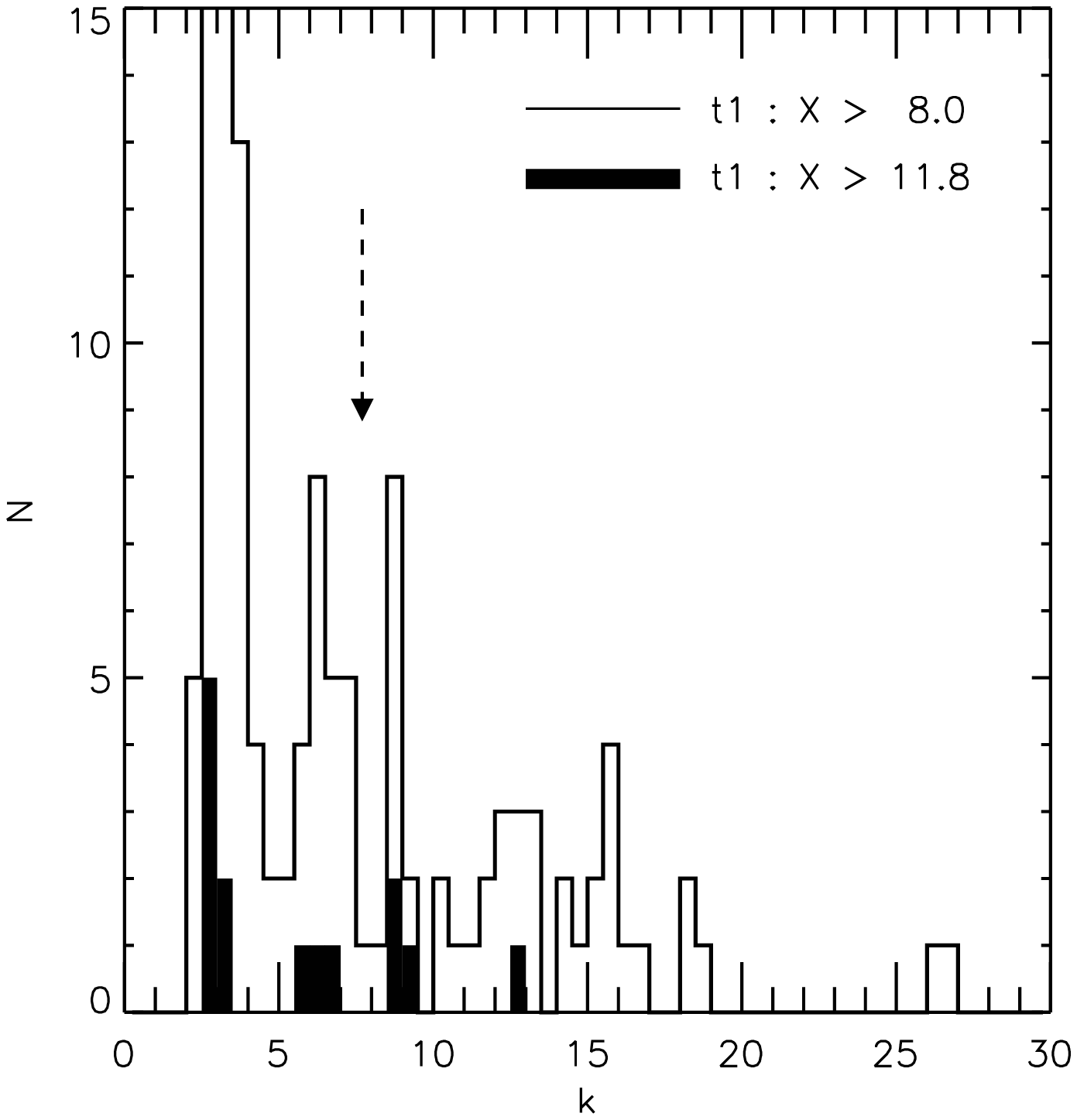,width=8cm,height=6cm}\\ 
\epsfig{file=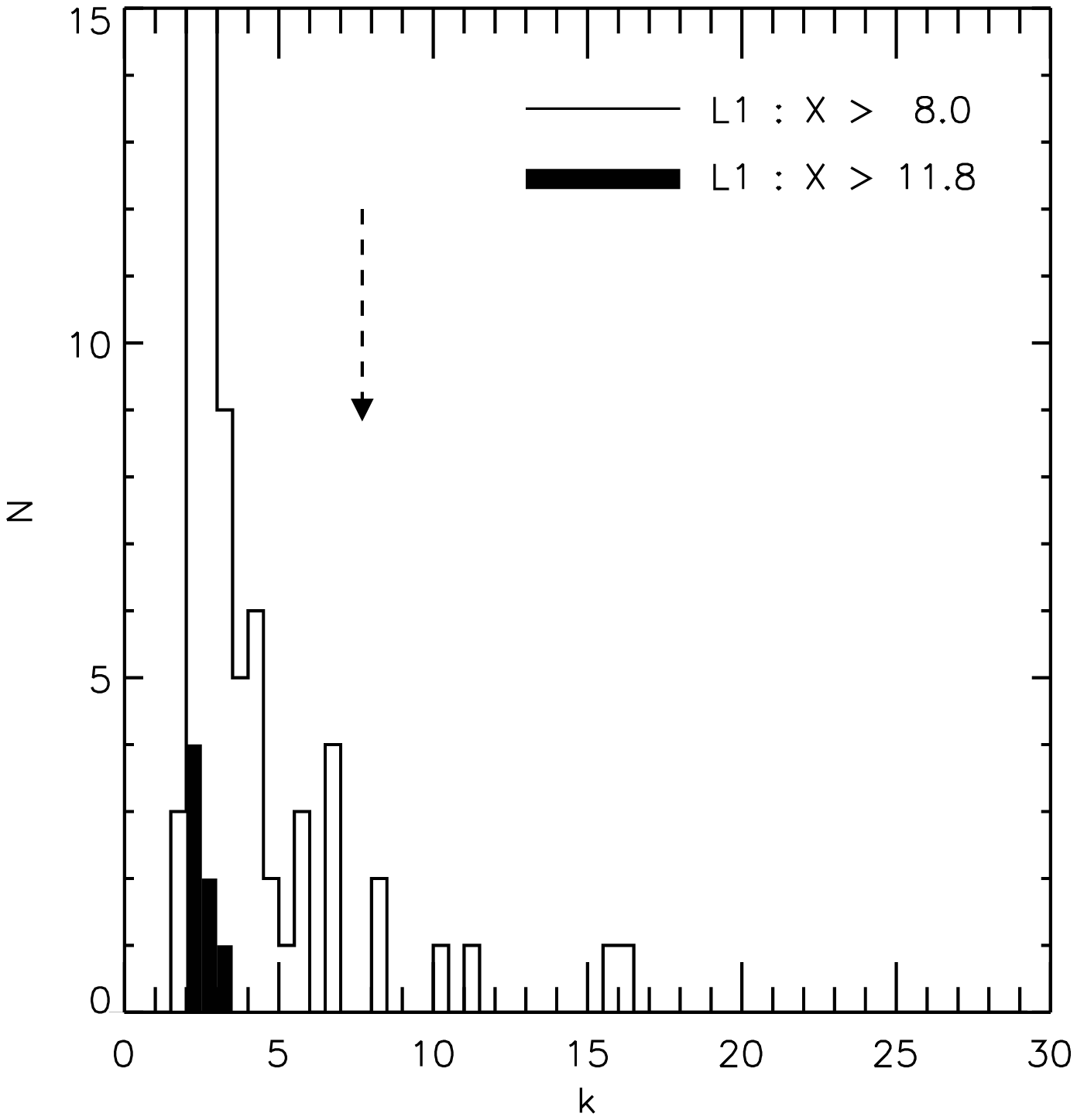,width=8cm,height=6cm}\\ 
\epsfig{file=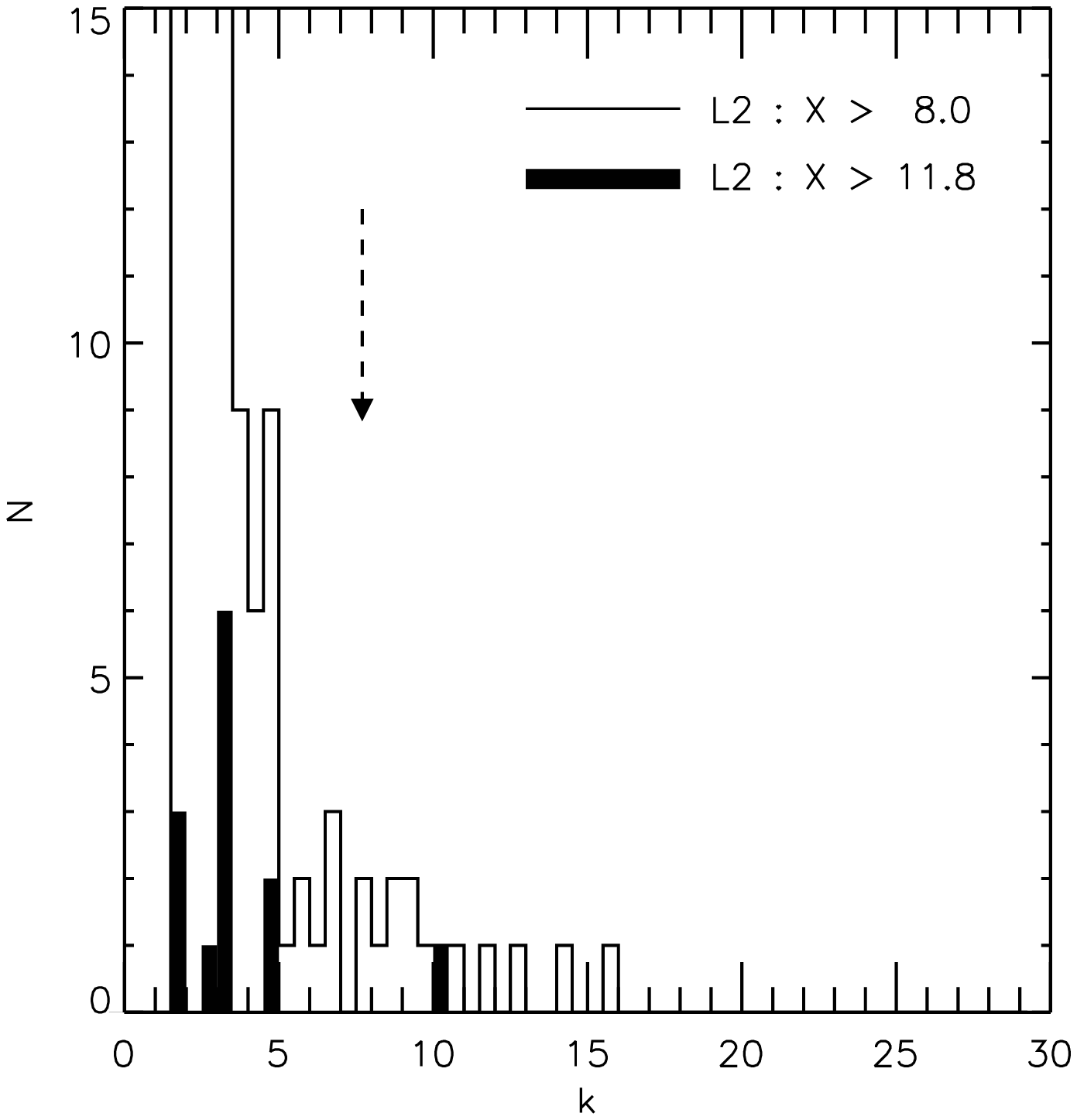,width=8cm,height=6cm} 
\caption{The distribution of wavenumbers for the high peaks. 
The shaded histograms show the number counts of the peaks with 
$X\ge X_{\rm BEKS}$, and the unshaded histograms show those with $X\ge 8.0$.
The arrow indicates the peak wavenumber of the BEKS data.}
\label{peaks}
\end{center}
\end{figure}

\begin{figure}
\begin{center}
\epsfig{file=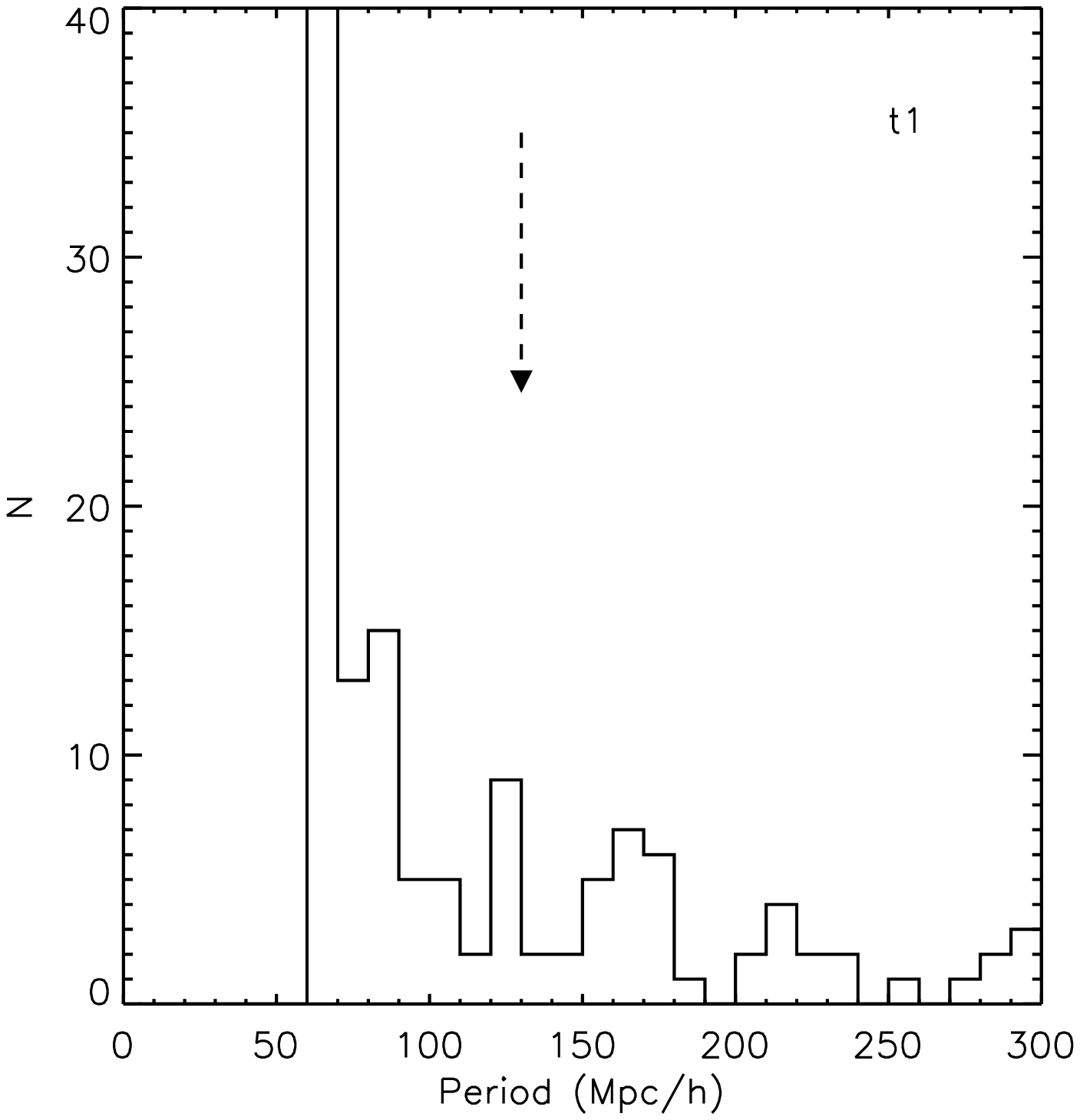,width=8.2cm,height=6.2cm}\\ 
\epsfig{file=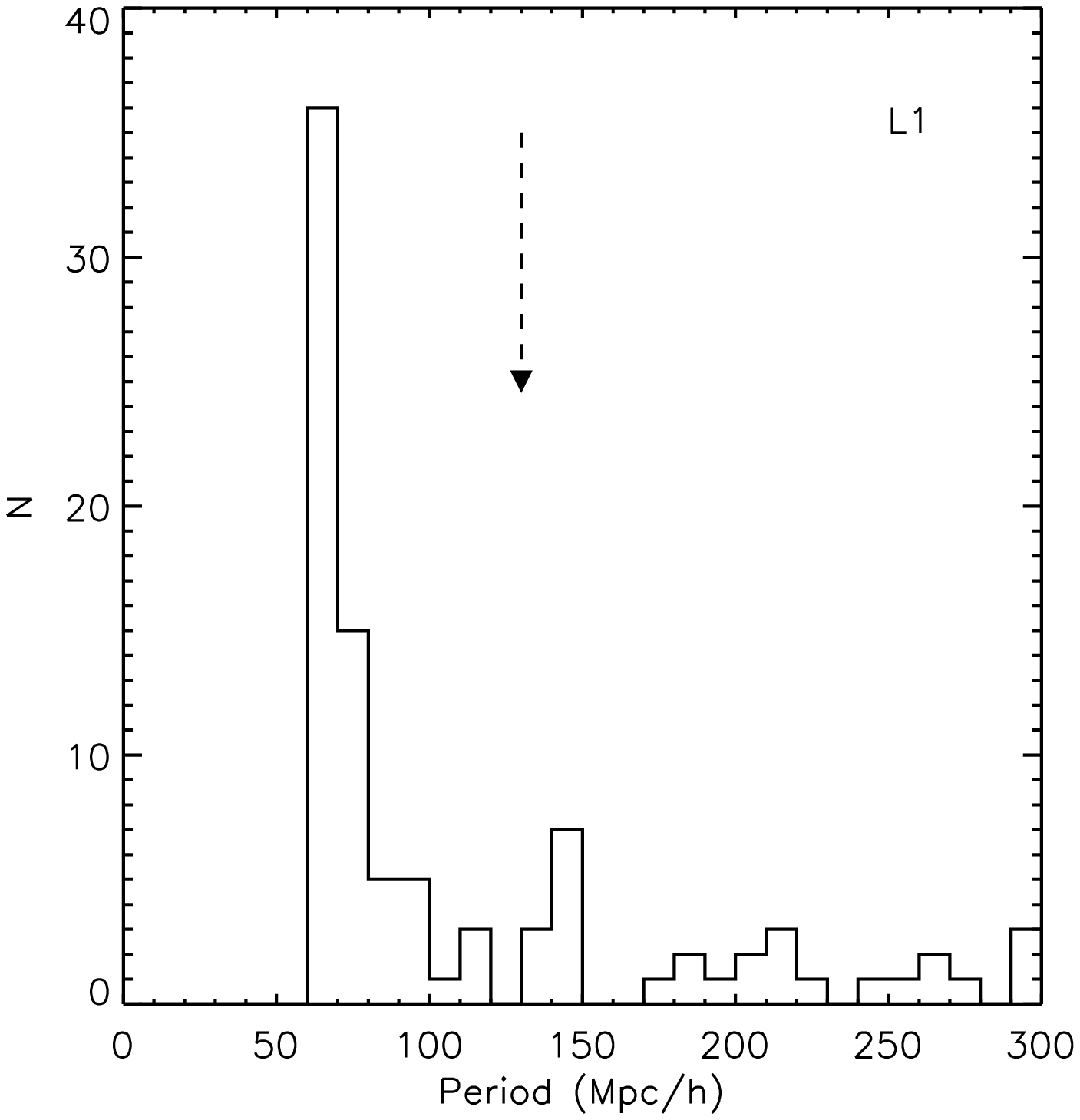,width=8.2cm,height=6.2cm}\\ 
\epsfig{file=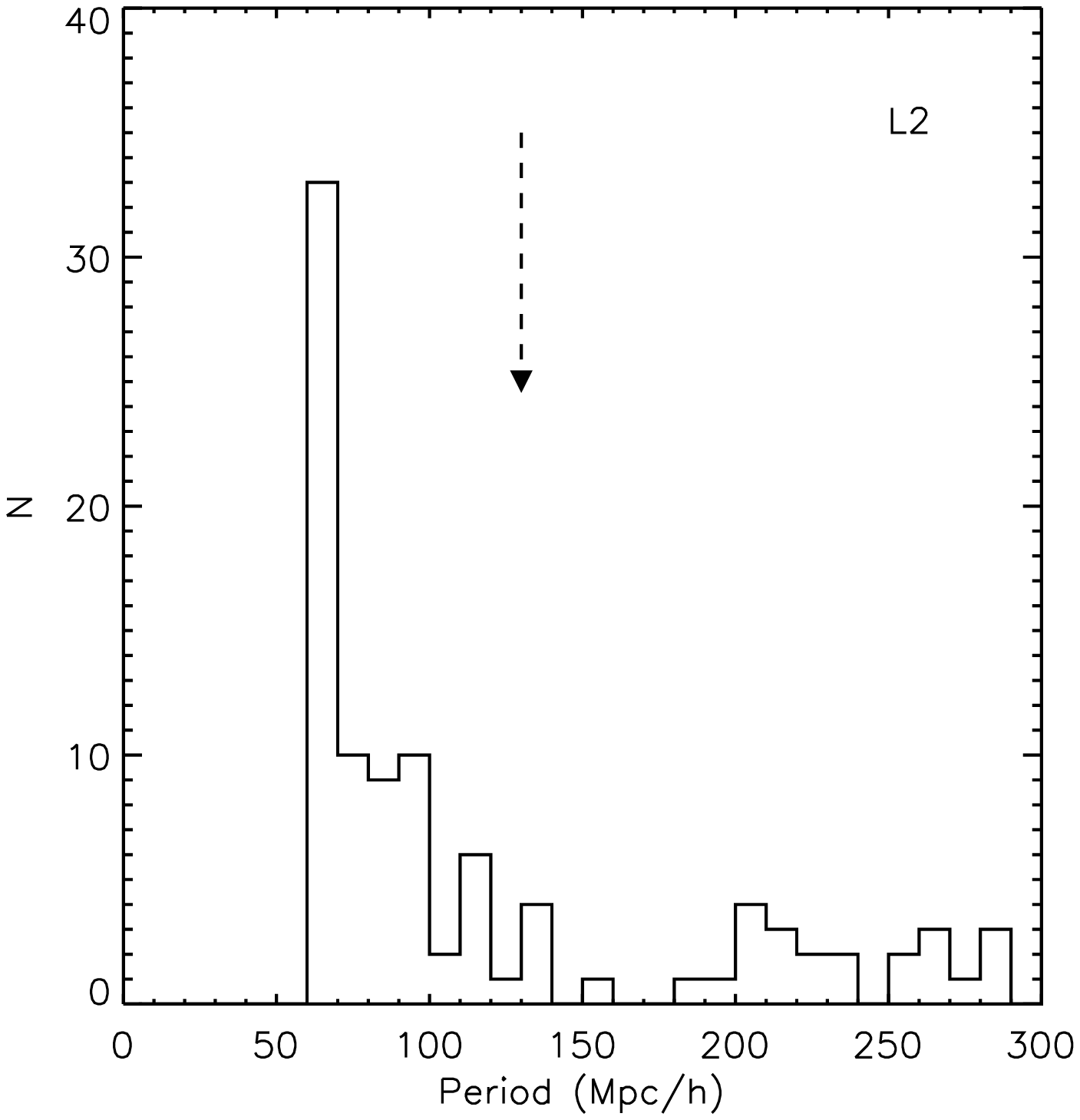,width=8.2cm,height=6.2cm} 
\caption{$\Delta$-test: the number counts of 
selected samples with $\Delta < \Delta$(BEKS) are plotted
against the measured period (see text).  
The period of the BEKS data is shown by an
arrow at 130 $h^{-1}$Mpc.}
\label{lowdelta}
\end{center}
\end{figure}

\begin{figure}
\begin{center}
\epsfig{file=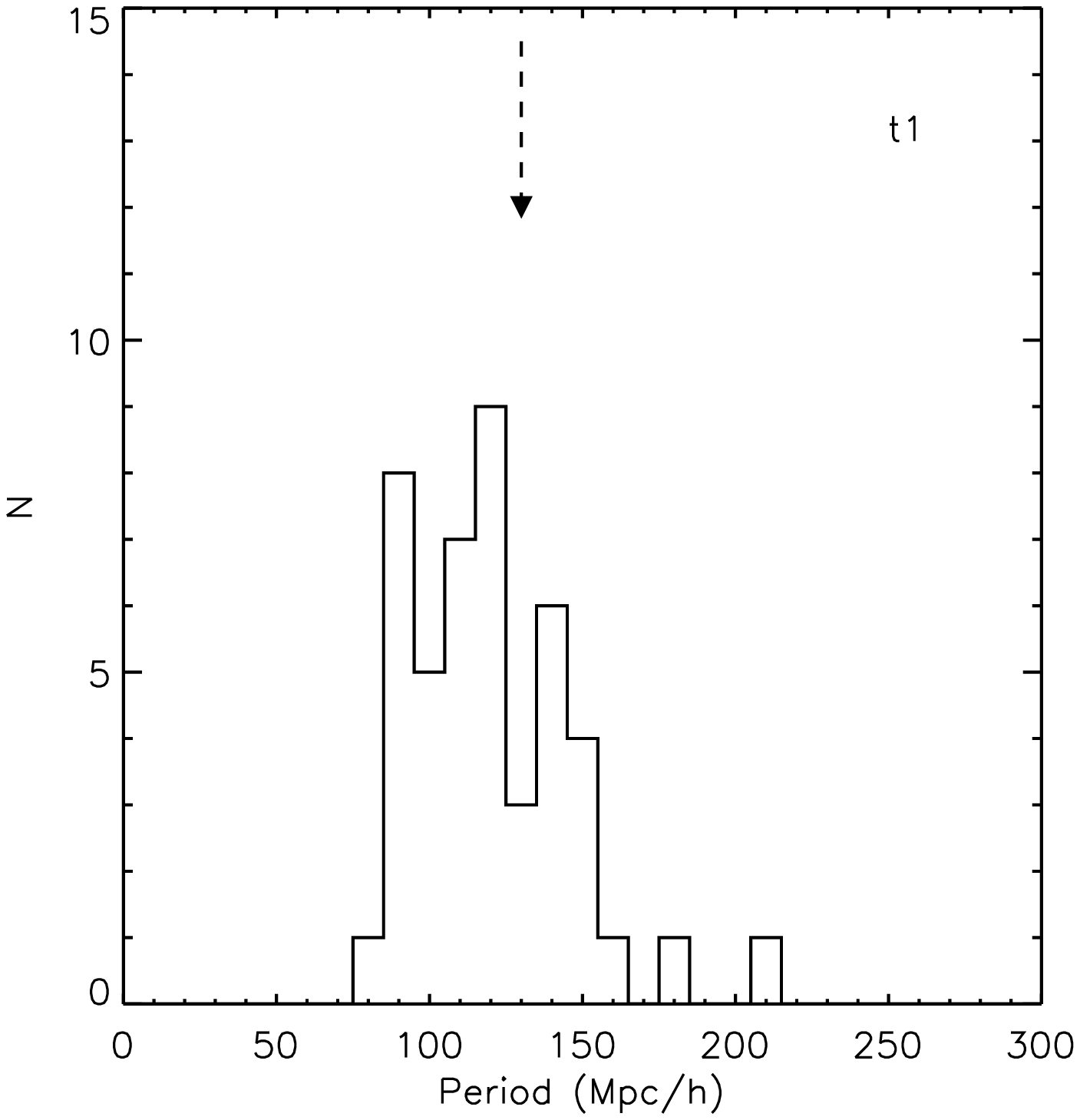,width=8.2cm,height=6.2cm}\\ 
\epsfig{file=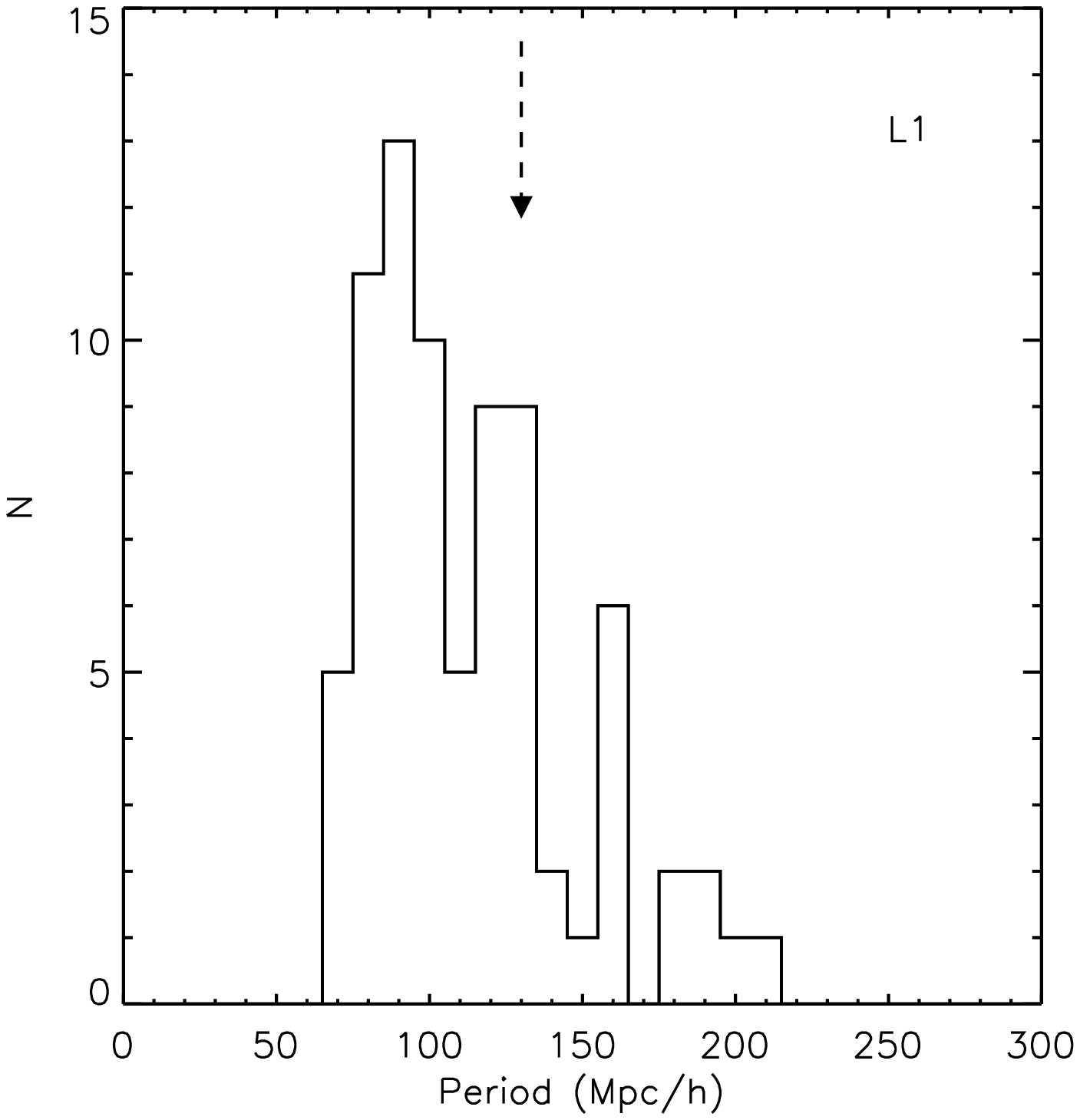,width=8.2cm,height=6.2cm}\\ 
\epsfig{file=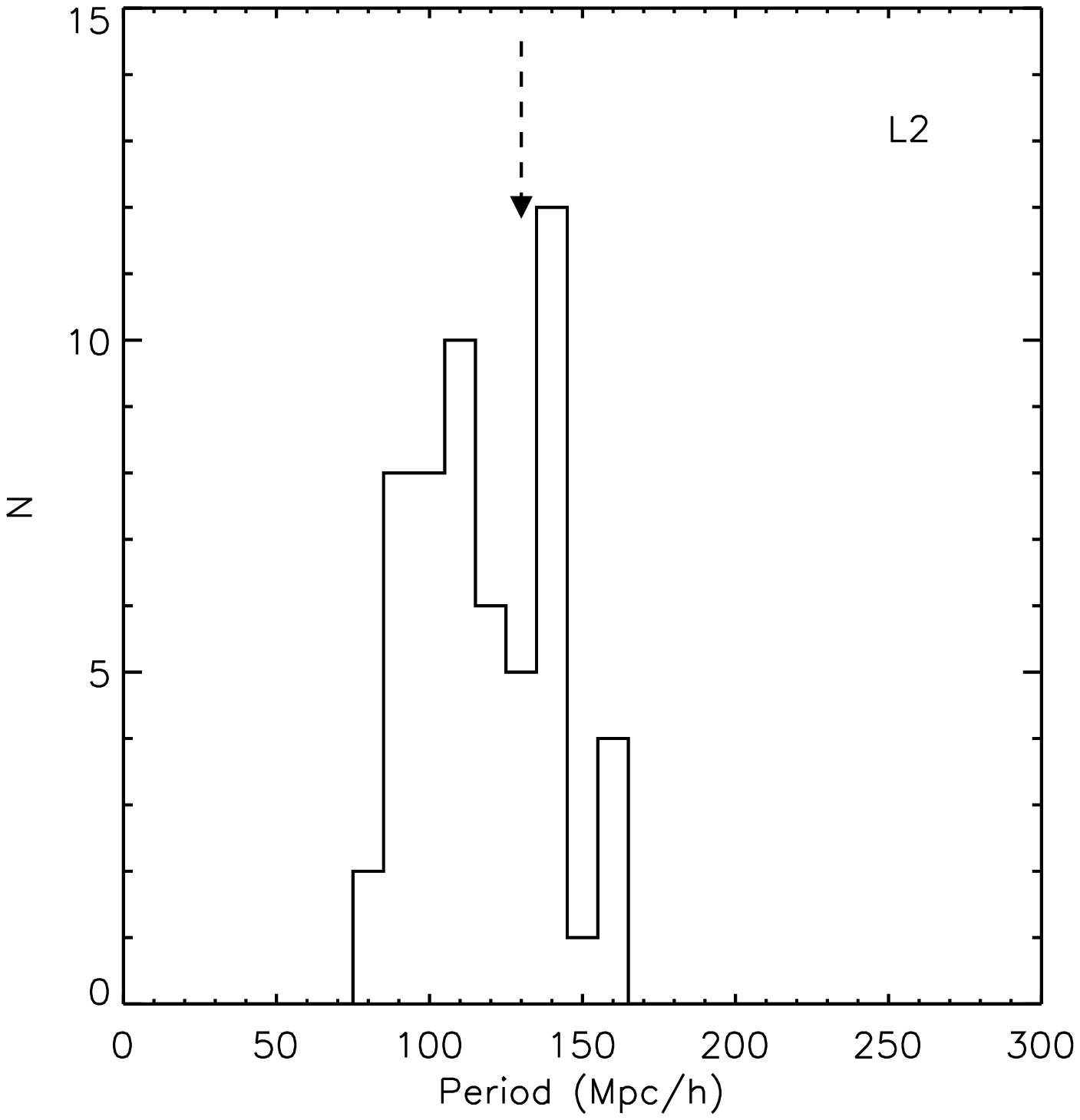,width=8.2cm,height=6.2cm} 
\caption{The Rayleigh statistic $R$. The number counts of $R < R(\mbox{BEKS})$,
plotted against the period assigned to each of the samples for 
the Rayleigh statistic. The arrow indicates the period for the BEKS data.}
\label{dist_R}
\end{center}
\end{figure}


\begin{thebibliography}{}
\bibitem[\protect\citename{Bahcall }1991]{Bah}%
Bahcall, N. A., 1991,
ApJ, 376, 43
\bibitem[\protect\citename{Baugh }1996]{B}%
Bahcall, C. M., 1996,
MNRAS, 280, 267
\bibitem[\protect\citename{Bellanger \& de Lapparent}1995]{BL}%
Bellanger, C. \& de Lapparent, V., 1995,
ApJ, 455, L103
\bibitem[\protect\citename{Benson et al.\ }2000]{Benson}%
Benson, A. J., Cole, S., Frenk, C. S., Baugh, C. M., Lacey, C. G., 2000, 
MNRAS, 311, 793
\bibitem[\protect\citename{Broadhurst et al.\ }1990]{BEKS}%
Broadhurst, T. J., Ellis, R. S., Koo, D. C. \& Szalay, A. S., 1990,
{\it Nature}, 343, 726
\bibitem[\protect\citename{Cohen\ }1999]{Cohen}%
Cohen, J. G., 1999,
in Mazure, A. \& Le F\`{e}vre, O. eds, 
Clustering at High Redshift, 
ASP Conference Series, Vol. 200, p314
\bibitem[\protect\citename{Cole et al.\ }1998]{CHF}%
Cole, S. M., Hatton, S. J., Weinberg, D. H., \& Frenck, C. S., 1998,
MNRAS, 300, 505
\bibitem[\protect\citename{Coles }1990]{Co}%
Coles, P. 1990,
{\it Nature},  346, 446 
\bibitem[\protect\citename{Dekel et al.\ }1992]{Dekel}%
Dekel, A., Blumenthal, G. R., Primack, J. R., \& Stanhill,D., 1992,
MNRAS, 257, 715 
\bibitem[\protect\citename{Dekel \& Lahav\ }1999]{DekelLahav}%
Dekel, A. \& Lahav, )., 1999, 
ApJ, 520, 24
\bibitem[\protect\citename{Duari et al.\ }1992]{Duari}%
Duari, D., Das Gupta, P., \& Narlikar, J. V., 1992, 
ApJ, 384, 35 
\bibitem[\protect\citename{Einasto et al.\ }1997]{Einasto}%
Einasto, M., Tago, E., Jaaniste, J., Einasto, J., \& Andernach, H,
1997,
Astron. Astrophys,  123, 119. 
\bibitem[\protect\citename{Eisenstein et al.\ }1998]{Eisenstein}%
Eisenstein, D. J., Hu, W., Silk, J., \& Szalay, A. S., 1998,
ApJ, 494, L1
\bibitem[\protect\citename{Evrard et al.\ }2000]{Hubble}%
Evrard, A.E., MacFarland, T.J., Couchman, H.M.P, Colberg, J.M.,
Yoshida, N., White, S.D.M., Jenkins, A., Frenk, C.S., Pearce, F.R.,
Efstathiou, G., Peacock, J.A., Thomas, P.A. 2000, 
in preparation 
\bibitem[\protect\citename{Ettori et al.\ }1999]{EGT}%
Ettori, S., Guzzo, L., \& Tarenghi, M., 1997,
MNRAS, 285, 218
\bibitem[\protect\citename{Feller }1999]{Feller}%
Feller, W., 1971,
{\it An Introduction to Probability Theory and Its Applications},
Volume 2, 2nd Edition, (New York: John Wiley and Sons)
\bibitem[\protect\citename{F\`{e}vre et al.\ }1999]{Fevre}%
Le F\`{e}vre, O., et al., 1999,
in Giuricin, G., Mezzetti, M., \& Salucci, P. eds, 
Observational Cosmology: The development of Galaxy Systems,
ASP Conference Series, Vol. 176, p250
\bibitem[\protect\citename{Frenk }1991]{Frenk}%
Frenk, C. S., 1991,
Physica Scripta, Vol. T36, p70
\bibitem[\protect\citename{Guzzo }1999]{Guzzo}%
Guzzo, L. 1998,
in Paul, J., et al. eds, Abstracts of the 19th Texas Symposium on Relativistic 
Astrophysics and Cosmology, CEA Saclay, p 510
\bibitem[\protect\citename{Hill, Steinhardt \& Turner }1991]{Hill}%
Hill, C. T., Steinhardt, P. J., \& Turner, M., 1991.
ApJ, 366, L57
\bibitem[\protect\citename{Jenkins et al.\ }1998]{virgo}%
Jenkins, A. et al., 1998,
ApJ, 499, 20
\bibitem[\protect\citename{Kaiser \& Peacock}1991]{KP}%
Kaiser, N. \& Peacock, J.A., 1991,
ApJ, 379, 482
\bibitem[\protect\citename{Kauffmann et al.\ }1999]{Kauffmann}%
Kauffmann, G., Colberg, J. M., Diaferio, A, \& White, S. D. M., 1999,
MNRAS, 303, 188
\bibitem[\protect\citename{Koo et al.\ }1993]{Koo}%
Koo, D.C., Ellmann, N., Kron, R.G., Munn, J.A., Szalay,
A.S., Broahurst, T.J., \& Ellis, R.S., 1993, 
in Chiancarini, G., Iovino, A., \& Maccagni, D. eds.,
Observational Cosmology
\bibitem[\protect\citename{Kurki }1990]{Kurki}%
Kurki-Suonio, H., Matthew, G.J., \& Fuller, G.M., 1990,
ApJ, 336, L5
\bibitem[\protect\citename{Morikawa }1990]{Morikawa}%
Morikawa, M., 1990,
ApJ, 362, L37
\bibitem[\protect\citename{Park \& Gott }1991]{PG}%
Park, C. \& Gott, J.R., 1991,
MNRAS, 249, 288
\bibitem[\protect\citename{Pierre }1990]{Pi}%
Pierre, M. 1990, 
A\&A, 229, 7
\bibitem[\protect\citename{Sigad et al.\ }2000]{Sig}%
Sigad, Y., Branchini, E., \& Dekel, A., 2000,
ApJ, 540, 62
\bibitem[\protect\citename{Somerville et al.\ }2001]{Som}%
Somerville, R.S., Lemson, G., Sigad, Y., Dekel, A., 
Kauffmann, G., \& White, S.D.M., 2001,
MNRAS, 320, 289
\bibitem[\protect\citename{Szalay et al.\ }1991]{SKEB}%
Szalay, A.S., Koo, D.C., Ellis, R.S., \& Broadhurst, T.J., 1991,
in S.Holt, C.Benett, \& V.Trimble eds., 
After the Furst Three Minutes, (New York:AIP), p261 
\bibitem[\protect\citename{SubbaRao \& Szalay }1992]{SS}%
SubbaRao, M.U., \& Szalay, A.S., 1992,
ApJ, 391, 483
\bibitem[\protect\citename{Tully et al.\ }1992]{TSVZ}%
Tully, R.B, Scaramella, R., Vettolani, G. \& Zamorani, G., 1992,
ApJ, 388, 9
\bibitem[\protect\citename{van de Weygaert }1991]{vd}%
van de Weygaert, R. 1991,
MNRAS, 249, 159
\bibitem[\protect\citename{Willmer et al.\ }1994]{WKSK}%
Willmer, C.N.A., Koo, D.C., Szalay, A.S. \& Kurtz, M.J., 1994,
ApJ, 437, 560 
\bibitem[\protect\citename{White et al.\ }1987]{WFDE}%
White, S.D.M., Frenk, C.S., Davis, M., \& Efstathiou, G., 1987,
ApJ, 313, 505
\end{thebibliography}
\end{document}